# Knowledge-driven deep learning for fast MR imaging: undersampled MR image reconstruction from supervised to un-supervised learning

*Shanshan Wang[1,2], Ruoyou Wu[1], Sen Jia[1], Alou Diakite[1,3], Cheng Li[1], Qiegen Liu[5], and Leslie Ying[4]*

[1] Paul C Lauterbur Research Center for Biomedical Imaging, Shenzhen Institute of Advanced Technology, Chinese Academy of Sciences
[2] Guangdong Provincial Key Laboratory of Artificial Intelligence in Medical Image Analysis and Application, Guangdong, China
[3] University of Chinese Academy of Sciences, Beijing, China
[4] Department of Biomedical Engineering and Department of Electrical Engineering, The State University of New York, Buffalo, New York 14260, USA
[5] Department of Electronic Information Engineering, Nanchang University, Nanchang, China

**Running title: Knowledge-driven deep learning for fast MR imaging**

**Word Count: ~ 8000**

Correspondence to **Shanshan Wang**, Ph.D.
Paul C. Lauterbur Research Centre for Biomedical Imaging
Institute of Biomedical and Health Engineering
Shenzhen Institute of Advanced Technology,
Chinese Academy of Sciences
Shenzhen, Guangdong, P. R. China, 518055
Tel: (86) 755 -8639224 3
Email: sophiasswang@hotmail.com

**Abstract:** Deep learning (DL) has emerged as a leading approach in accelerating MR imaging. It employs deep neural networks to extract knowledge from available datasets and then applies the trained networks to reconstruct accurate images from limited measurements. Unlike natural image restoration problems, MR imaging involves physics-based imaging processes, unique data properties, and diverse imaging tasks. This domain knowledge needs to be integrated with data-driven approaches. Our review will introduce the significant challenges faced by such knowledge-driven DL approaches in the context of fast MR imaging along with several notable solutions, which include learning neural networks and addressing different imaging application scenarios. The traits and trends of these techniques have also been given which have shifted from supervised learning to semi-supervised learning, and finally, to unsupervised learning methods. In addition, MR vendors' choices of DL reconstruction have been provided along with some discussions on open questions and future directions, which are critical for the reliable imaging systems.

## 1. Introduction

Magnetic Resonance Imaging (MRI) is a non-invasive medical imaging technology that can reveal different properties of underlying anatomical structures. Nevertheless, its data acquisition time is relatively too long. Long acquisition time dramatically limits the achievable spatial and temporal resolutions of MRI, and frequently leads to motion artifacts. These shortcomings make MRI exams inefficient, expensive, and sometimes infeasible for clinical applications where high temporal or spatial resolution imaging are needed.

Driven by the success of deep learning (DL) in computer vision (CV) and natural image processing (image restoration, denoising, etc.), DL based image reconstruction methods have gained popularity for accelerating MRI acquisition [1–5]. Neural networks are firstly trained to extract knowledge from available datasets and then are utilized to assist in image reconstruction from incomplete measurements. Comparing with conventional reconstruction methods for accelerating MR imaging, these DL based approaches have demonstrated improved reconstruction quality and provide great potential to achieve faster MR imaging [6]. However, different from natural images, MRI involves specific physics-based imaging processes, special properties of underlying data, and different imaging tasks or applications. Pure data-driven methods or the direct application of CV networks to MRI may suffer from several issues, such as poor generalization capability, unstable performance, poor expandability, etc. To address these issues, it is of great necessity to develop knowledge-driven deep learning techniques, which incorporate domain-specific knowledge such as imaging physics and prior knowledges about MRI into DL architecture designing and/or training.

This review will discuss the significant challenges faced by such knowledge-driven DL approaches in the context of fast MR imaging along with several notable solutions. It covers a broad spectrum, from neural network learning to diverse imaging application scenarios. We will also highlight the characteristics and recent trends of these techniques, which have experienced a paradigm shift from "big data, minimal physics" to "limited data, maximal physics". More specifically, the DL reconstruction technique developments are experiencing the transfer from fully supervised learning to unsupervised learning. This is the key difference between our review and the previous ones [6–9], the techniques which surveyed are mainly fully supervised, with limited attention to semi-supervised or unsupervised methods. A key insight of our review is trying to highlight that "domain knowledge" should be incorporated into deep learning that can address the challenges lying in the supervised methods for real-world MRI applications. Furthermore, different from [6–9] that review academic developments, we also review some MR vendors' choices of deep learning powered image reconstruction techniques and the special features of their commercial products. The differences and gaps in the development of DL MR reconstruction methods in industry and academy are also revealed. The deep learning reconstruction methods in the industry primarily rely on supervised learning, combining the physical models and iterative algorithms to provide interpretability. In contrast, the academic realm tends to explore unsupervised learning methods to reduce the dependence on fully sampled data. The technological developments applying DL for fast MRI have shown impressive performance in addressing the long-standing technical challenges associated with different applications scenario[10–12]. However, the scope of this review is undersampled MR image reconstruction with deep learning and the application scenarios mainly consist of contrast, dynamic and quantitative imaging. In addition, we have also provided three main types of deep learning methods for fast MRI imaging, along with open-source datasets and code links in Table S1. The methodologies used for the systematic review, including literature search, article screening, full-text reading and categorization are described in the appendix.

**2. Deep Learning Basics**

Deep learning (DL) is a branch of machine learning that allows computational models consisting of multiple processing layers to learn data representations with multiple levels of abstraction[13]. Unlike traditional approaches, deep learning enables computers to acquire knowledge from experience, eliminating the requirement for explicit human input to define all the necessary knowledge. By hierarchically constructing complex concepts from simpler ones, computers can grasp intricate ideas. Visualizing this hierarchy would reveal multiple layers, demonstrating the depth of the learning process. Based on the availability of labeled data, deep learning can be categorized into three major types: supervised learning, semi-supervised learning, and unsupervised learning.

**2.1. Supervised Learning**

Supervised learning is a machine learning paradigm in which a model learns from labeled training data [14]. Each training example consists of input features and corresponding output labels. The objective is to learn a function that can map new, unseen inputs to their correct output labels. This is achieved through optimization algorithms that minimize the discrepancy between predicted and actual labels. By learning the relationship between the input features and target variables, the algorithm builds a predictive model. When given new input data, the model can predict its corresponding target variable.

Mathematically, in supervised learning, we have a training dataset consisting of pairs of inputs $m_i$ and corresponding labels $v_i$, where $i$ denotes the index of the data point. We aim to learn a function $f(m)$ that maps the input space to the output space, such that $f(m_i)$ approximates $v_i$ for all training examples. This is achieved by minimizing an appropriate loss or error function, typically represented as:

$$\min_{\theta} \sum_i L(f(m_i), v_i, \theta) \qquad (1)$$

where $\theta$ denotes the parameters or weights of the model, and $L(\cdot)$ quantifies the discrepancy between the predicted output $f(m_i)$ and the true label $v_i$. The optimization process involves adjusting the weights iteratively using techniques such as gradient descent, aiming to find the optimal set of weights that minimizes the overall loss. In MR image reconstruction, supervised learning methods train a nonlinear mapping or draw knowledge between the undersampled and fully-sampled data pairs.

**2.2. Semi-Supervised Learning**

Semi-supervised learning is a machine learning method that combines labeled and unlabeled data for training. In semi-supervised learning, the algorithm uses labeled data to learn the model and leverages the unlabeled data to provide additional information, improving the model's performance and generalization. Semi-supervised learning is often applied in scenarios where labeled data is scarce or expensive to obtain, allowing the utilization of the potential information in unlabeled data [15].

Mathematically, in semi-supervised learning, we have a dataset that includes both labeled examples $\{(m_i, v_i)\}$ and unlabeled examples $m_j\}$. The objective is to learn a model that utilizes both labeled and unlabeled data to improve learning performance. This can be achieved by optimizing a combined loss function that incorporates both labeled and unlabeled data, such as:

$$\min_{\theta} \sum_i L(f(m_i), v_i, \theta) + \sum_j U(f(m_j), \theta) \qquad (2)$$

where $L(\cdot)$ represents the loss function for labeled data, $U(\cdot)$ represents the unsupervised loss function for unlabeled data, and $\theta$ denotes the model parameters. In MRI, both undersampled and fully-sampled data can be explored with the semi-supervised learning technique.

**2.3. Unsupervised Learning**

Unsupervised learning is a type of machine learning that looks for previously undetected patterns in a dataset without predefined output labels and with minimal human supervision. In unsupervised learning, the algorithm explores the features and relationships inherent in the data and discovers patterns and structures through techniques such as clustering, dimensionality reduction, and association rules [16].

Self-supervised learning is a variant of unsupervised learning where the model learns representations or features from the data itself using pretext tasks. These pretext tasks involve generating supervisory signals from the data, allowing the model to learn useful representations that can later be transferred to downstream tasks [16].

Mathematically, in unsupervised learning, we aim to learn a representation or structure of the data without access to explicit labels. This is typically achieved by optimizing a generative model that captures the underlying distribution of the data. For instance, let $M = \{m_1, m_2, \cdots, m_N\}$ denote a set of unlabeled data points. We seek to learn a model $p(m)$ that approximates the true data distribution $p_{data}(m)$. This can be accomplished by maximizing the likelihood of the data under the generative model, which can be written as:

$$\max_{\theta} \sum_i \log p(m_i, \theta) \qquad (3)$$

where $\theta$ represents the parameters of the generative model. The optimization process involves adjusting the model parameters to maximize the likelihood of the data. In MR image reconstruction, only undersampled k-space data are used to train the neural networks without the fully-sampled ground-truth data.

**3. Fast MR imaging basics and special properties**

The success of DL in CV and the natural image processing fields provides a new perspective to fast MRI approaches and give the potential to design more powerful acceleration methods [6,17,18]. However, MR images hold distinct properties that are significantly different from natural images, and the data acquisition and image reconstruction involve multiple physics and signal processing related processes. Therefore, developing DL based acceleration techniques must comprehensively encompass the domain specific knowledge such as the characteristics of MR images, the physics underlying MR data acquisition and the intended clinical imaging applications.

## 3.1. Physics of MR Imaging

The imaging process could be described by the phenomenological Bloch equation [19]. Solving the Bloch equation for the interaction of the spin magnetization with the spatial encoding gradients derives the MR signal equation which establishes the Fourier Transform (FT) relationship between the measured signal $m_j(\vec{k})$ in k-space and the transverse spin magnetization $v(\vec{x})$ in the spatial domain:

$$m_j(\vec{k}) = \int v(\vec{x}) \cdot e^{i\varphi(\vec{x})} \cdot C_j(\vec{x}) \cdot e^{-i\vec{k}\vec{x}} d\vec{x}, \quad \vec{k} = \frac{\gamma}{2\pi}\int_0^t \vec{G}(\tau)d\tau \quad (4)$$

where $\vec{x}$ denotes the spatial position of voxels in the imaged volume, and $\vec{k}$ denotes the spatial frequency in k-space. $C_j(\vec{x})$ denotes the coil sensitivity of the multi-channel parallel receiving coil, $\varphi(\vec{x})$ represents the spatially varying phase information in MRI. According to the Nyquist sampling theorem, the MR signal equation mentioned above can be discretized into a linear system of equations as follows:

$$KFCv = m \quad (5)$$

where $v$ and $m$ denote the complex-valued vectors containing the unknown discrete image pixels and the acquired k-space samples, $C$ represents the coil sensitivity encoding operator, $F$ represents the Fourier encoding operator of the gradient field system, and $K$ denotes the discrete sampling operator.

The Fourier Transform based spatial encoding in k-space of MRI is achieved by repetitively switching on and off the magnetic field gradients, leading MR data acquisition to be quite slow, and significantly limiting the achievable spatial and temporal resolution of MRI. Additionally, there are some other limiting factors such as contrast encoding, the inefficiency of spin-warp Fourier encoding, constraints on gradient slew and amplitude, physiologic constraints, and SNR limitations. More knowledges on MR physics and background can be referred to [20].

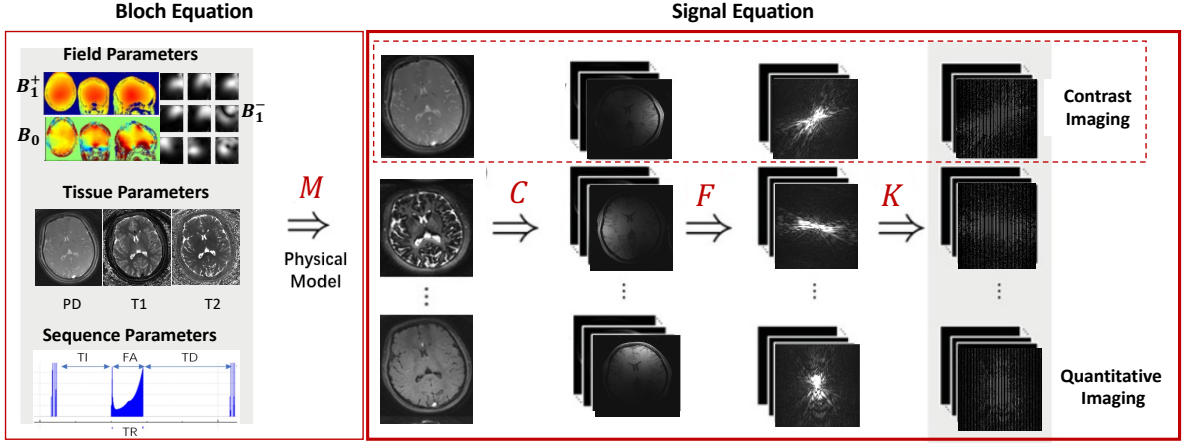

Figure 1

## 3.2. Accelerating MR data acquisition by undersampling

Currently, reducing the number of k-space samples is the most important way to accelerate MR imaging. The missing data needs to be estimated by designing sophisticated image reconstruction algorithms based on the knowledge of MR images, k-space samples, and the imaging model. Partial Fourier (PF), Parallel Imaging (PI), and Compressed Sensing (CS) are three widely used fast MR imaging techniques. PF utilizes the knowledge that the phase of a complex-valued MR image may vary smoothly, enabling less than half k-space undersampling and smooth phase constrained image reconstruction. PI utilizes the spatial encoding capabilities of multiple coil sensitivities to design regular undersampling and linear inverse-based image reconstruction. Based on the sparsity knowledge of MR images, CS designs incoherent k-space undersampling and nonlinear inverse reconstruction to achieve higher acceleration. The image reconstruction models for these fast MR imaging techniques can be unified as the following optimization problem:

$$\hat{v} = \min_{v} \ \|KFCv - m\|_2^2 + \sum_i \lambda_i \mathcal{R}_i(v) \tag{6}$$

where $\|KFCv - m\|_2^2$ enforces the acquired data consistency, $\mathcal{R}_i(\cdot)$ denotes the different regularization terms used in the fast MR imaging techniques, and $\lambda_i$ denotes the regularization weights. The optimization problem in Eq.6 can be solved by employing different iterative algorithms according to the regularization term used, e.g., linear conjugate gradient (CG) algorithm when $\mathcal{R}(v) = \|v\|_2^2$ or nonlinear fast iterative shrinkage thresholding (FISTA) algorithm when $\mathcal{R}(v) = \|Wv\|_1$. **Figure 1** presents several variations of Eq.6 in commonly used fast MR imaging scenarios.

Each of the fast MR imaging techniques have their own limitations and challenges. The reconstruction quality of PF heavily relies on the accuracy of phase estimation. PI mainly faces the challenge of significant noise amplification at high acceleration factors. The PI reconstruction quality depends on the coil configuration, the accuracy of PI model, and the choice of sampling patterns. CS requires empirically tuning the sparsity regularization weight to balance between the noise suppression and the fidelity of image details. It faces challenges associated with extended reconstruction time due to the increased computational complexity of nonlinear reconstruction algorithm [1–5]. To address these challenges, deep learning can be incorporated into the whole MR imaging pipeline as shown in **Figure 2**, which involves multiple signal processing steps. The pre-processing steps work on the raw data to conduct noise pre-whitening, remove readout oversampling and coil compression, etc. Post-processing steps involve manipulating the reconstructed image to perform tasks like filtering, interpolation, inhomogeneity correction and data type conversion (from complex flat to integer). But the scope of this review is undersampled MR image reconstruction with deep learning and the application scenarios mainly consist of contrast, dynamic and quantitative imaging.

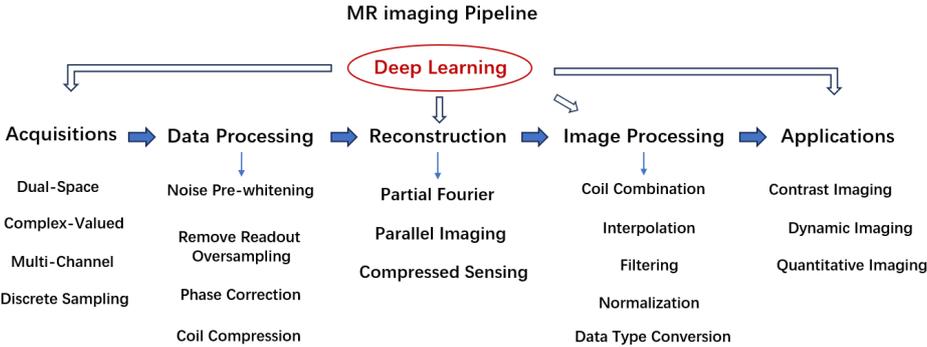

Figure 2

## 3.3. Fourier Imaging-Dual Space Property

The MRI raw data is acquired in the spatial frequency domain, also known as k-space, which represents the discrete Fourier transform of the unknown MR images. The image reconstruction needs to estimate an artifact-free image of high signal-to-noise ratio (SNR) from undersampled k-space samples. Deep learning is a powerful nonlinear mapping tool to learn/extract important features or prior knowledge from available dataset. Different from natural image processing that deep learning mainly operates in the spatial domain, deep learning for fast MR imaging can either directly learn the nonlinear mapping between the aliased MR images and the reference MR images, or between the undersampled k-space data and fully sampled k-space data, or conduct cross-domain dual-space learning [21-29]. The dual-space property of MRI provides opportunities for DL reconstruction methods operating flexibly in image space, k-space, or dual space. There are several publicly available MR image databases for tasks related to DL based MR image processing or analysis. Thus, at the early stage of developing DL reconstruction methods, synthetic k-space data via applying an inverse Fourier transform to these images which frequently in the DICOM format are utilized for network training. However, DL reconstruction network trained by these simplistic synthetic k-space data would lead to implicit "data crimes" and systematically biased results when testing as pointed in [30]. Then, some publicly available databases have provided real-world full-sampled MR k-space raw data of several specific organs such as knee, brain, and cardiac (please see **Table S1 in the support information** regarding the available datasets and codes). Researchers can conduct k-space learning to directly interpolate the missing points [31] and have achieved encouraging results. It is empirically found out that deep learning in spatial domain is good at removing noise, but learning in the k-space domain possesses strong capability in capturing details. Therefore, Deep learning in both spatial domain and k-space domain are often recommended [21–23].

### 3.4. Complex-Valued MR Images

In typical MR imaging scenarios, spatially varying background phase always exists in the MR images, induced by factors like field inhomogeneity, magnetic susceptibility, chemical shift, or fluid flow, etc. [32]. As a result, both MR images and raw data are inherently complex-valued. The phase of MR images can also be manually controlled by designing RF pulses or gradients to provide several phase contrast MRI applications such as susceptibility weighted, angiography, flow and chemical shift imaging, etc. Hence, DL reconstruction methods need to be capable of handling the complex-valued data and computations, such as convolutions, activation functions, and loss functions.

However, the initial DL frameworks developed for natural image processing tasks only support real-valued inputs and computations. Consequently, most DL reconstruction methods split the real and imaginary components of the complex-valued MR dataset into two separate real-valued channels. For instance, Lee et al. [33] developed a computational-efficient MR image reconstruction model using deep residual learning networks with separated magnitude and phase networks. Feng et al. [34] proposed a dual-octave network (DONet) to jointly reconstruct the real and imaginary components of MRI as well as different spatial frequency groups. Xiao et al. [35] proposed a complex-valued CNN with an unrolled network for partial Fourier reconstruction. Cole et al. [36] analyzed the importance of used complex-valued operations in CNNs for complex-valued MRI reconstruction. Several recent works have also demonstrated the advantages and significance of building complex-valued networks for DL MR reconstruction [35,37–39]. It is desirable for DL networks to be able to operate on complex-valued data. And it is worth noting that the nonlinear activation function can be complex-valued as well [36].

**3.5. Parallel Imaging**

Most clinical MR scanners are equipped with multi-channel coils for signal reception and use PI reconstruction methods to accelerate MR scans of diverse organs. The key to PI reconstruction is accurately calibrating the coil sensitivity. Previously, two distinct categories of PI reconstruction model were prevalent: one relied on explicit estimation of coil sensitivities in the image domain, such as SENSE [40] and the other involved estimating a convolution kernel that implicitly represents coil sensitivities in k-space, such as GRAPPA [41] and SPIRiT [42]. Recently, subspace-based coil sensitivities calibration methods such as PRUNO [43], ESPIRiT [44], and SNMs [45], bridge the gap between the two categories of PI model, further enabling some calibrationless PI reconstruction models such as SAKE [46], LORAKS [47] and ALOHA [48].

Early DL reconstruction methods utilized single coil data for algorithm prototyping. Subsequently, DL reconstruction methods for multi-coil data were developed [49,50]. Most are based on the image domain SENSE model for coil combination or for building the data consistency (DC) term [51–53]. Some methods attempt to utilize the intrinsic relationship between coils using neural networks, without explicitly estimating the sensitivity maps of the coils [54-59]. Other methods attempt to simultaneously estimate coil sensitivity maps and perform image reconstruction [51,53,60]. These methods generally require a larger number of parameters, but they tend to achieve relatively better reconstruction quality.

Parallel imaging in k-space helps reduce the interference of noise. Traditional k-space based PI methods utilize autocalibration (ACS) signals from multiple coils to estimate the missing signals in the surrounding area. This mapping relationship is generally linear [41]. Leveraging the nonlinear modeling capabilities of deep neural networks, some researchers have started applying deep learning to k-space based PI [61-65]. Compared to traditional linear interpolation methods, the performance of nonlinear interpolation methods with the assistance of deep neural networks is superior, and the models are more robust [61-65]. More information on deep learning for parallel MR imaging can be referred to this review [7].

**3.6. Cartesian and Non-Cartesian Undersampling**

Most clinical MRI sequences acquire k-space samples using a Cartesian trajectory. The transformation between k-space and image can be computed fast using fast Fourier transform. Conventional fast MRI techniques often adopt different undersampling strategies, for example CAIPIRINHA undersampling for PI [66], variable-density Poisson-disc undersampling for CS [67] or inline adaptive sampling to achieve optimal reconstruction quality [68,69]. Sampling along a non-Cartesian trajectory, e.g., radial, spiral, or rosette trajectory, also provides several clinical benefits, such as sampling densely in the k-space center to follow the image contrast variations throughout the imaging process. Non-uniform FFT and accurate knowledge of the trajectory are required to transform non-Cartesian k-space samples to images, leading to increased computational burden and complexity. Non-Cartesian undersampling also offers several benefits such as lower noise enhancements and fewer coherent artifacts over Cartesian undersampling for subsequent PI and CS reconstruction [70].

Given the advantages of non-Cartesian sampling, research on reconstruction using non-Cartesian sampling patterns has gradually gained attention [71-73,75]. These approaches are mostly model-based, which enhances the interpretability of the methods. For example, Malavé et al. [71] built the reconstruction model by unfolding the proximal gradient descent (PGD) algorithm. By incorporating a non-uniform Fast Fourier Transform (NUFFT) operator, their model-based approach could be readily applied to reconstruct non-Cartesian data. NUFFT can be used in deep learning methods for non-Cartesian sampling. It converts non-uniformly sampled k-space data into images. However, its drawback is that the computation can be relatively slow. Additionally, density compensation methods have been introduced into non-Cartesian sampling patterns [74,76]. This allows for the adjustment of data sampling density in non-uniform sampling patterns and facilitates more accurate reconstruction of the original image by adjusting the weight of each sampled point. It aids in reducing artifacts, thereby improving image quality. Such as, Ramzi et al. [76] introduced the first density-compensated deep neural network for non-Cartesian k-space data and achieved better reconstruction performance than existing methods. There have also been methods trying to optimize sampling trajectories and image reconstruction jointly [69,77-84]. For example, Radhakrishna et al. proposed to jointly learn non-Cartesian k-space sampling trajectories and image reconstruction networks for 2D and 3D MRI [79]. With a VarNet, Sherry et al. [83] learned the optimal sampling pattern that can balance the acquisition time and reconstructed image quality. These methods further enhance the quality of dynamic imaging, but the computational burden increases accordingly.

Although DL reconstruction methods have been developed for both Cartesian and non-Cartesian trajectories [85] with various undersampling patterns, the inherent nonlinearity of DL reconstruction poses challenges in prospectively assessing the impact of diverse undersampling patterns and trajectories on reconstruction quality. To date, in most cases, the selection of trajectory and undersampling strategies for DL reconstruction are often made empirically.

## 3.7. Multi-Contrast Imaging

During a clinical MR exam, multi-contrast images (e.g., T1w, T2w, FLAIR, DWI, contrast-enhancement, etc.) are acquired sequentially at the same spatial position of lesions for complementary diagnostic information, which prolongs the scan time. These contrast weighted images may be acquired by multiple scans using different sequences, echo types, sequence timing, and magnetization preparation schemes, therefore leading to significantly different image contents and SNR. Moreover, the image contrasts may vary with the magnetic field strengths, MR vendors' implementations of a specific sequence, or even with the difference subjects.

Traditional multi-contrast imaging methods typically require manual adjustment of parameters or multi-step processing to integrate information from different contrasts, relying on domain expertise. In contrast, deep learning-based multi-contrast imaging methods are typically more flexible and adaptable to different tasks and data variations. They can undergo end-to-end training with large datasets, providing greater flexibility. Consequently, they may require more computational resources [86,91]. There have been some efforts to investigate deep learning for multi-contrast MR imaging [87-95]. For example, Kim et al. [87] improved the quality of MR image reconstruction by incorporating the information from high-resolution images of another contrast using adversarial learning. Do et al. [89] designed two networks, X-net and Y-net, to reconstruct multi-contrast MR images, where X-net reconstructs both T1- and T2-weighted images and Y-net reconstructs T2-weigthed images from downsampled T2-weighted images and fully sampled T1-weighted images. Additionally, recently popular diffusion models have provided a promising approach for MR contrast synthesis [96].

The flexible and abundant image contrasts provided by MRI arise several challenges to the development of supervised DL reconstruction techniques. Firstly, constructing a database of fully sampled multi-contrast k-space raw data is very prohibitively expensive or even infeasible. Secondly, the deviations between training and test data, for example using non-contrast enhanced data from healthy volunteer for training while contrast enhanced data from patients for testing, may degrade the testing image reconstruction quality.

### 3.8. Time-Resolved Dynamic Imaging

Dynamic MRI aims to imaging the dynamic processes occurring within the human body, encompassing both endogenous and exogenous physiological motions like heartbeat, flow, and dynamic contrast enhancement (DCE), as well as the evolution of physical signals such as tissue relaxation. However, the slow data acquisition speed of MRI leads to an intense tradeoff between the spatial and time resolution of dynamic MRI. Consequently, undersampling becomes essential to obtain clinically meaningful dynamic images. Fortunately, the additional data dimension (i.e., time) providing spatiotemporal correlation within dynamic images can be utilized to design powerful low-dimensional image representations, for example, low rank matrix or tensor, low-dimensional subspace, etc. These prior knowledges enable highly accelerated undersampling to effectively alleviate the tradeoff between spatial and time resolution.

Traditional methods typically employ physical or mathematical models to describe the evolution of dynamic MR signals, often involving frame-by frame reconstruction of dynamic image sequences. In contrast, deep learning methods can utilize neural networks to automatically extract features for modeling dynamic information. They can learn the representation of dynamic image directly from raw data through end-to-end learning, facilitating the joint reconstruction of the entire sequence. In dynamic MRI, deep learning methods have been widely applied [97-122]. They can be broadly categorized into methods based on spatiotemporal consistency [99,100,114,115] and convolution neural networks [97,111-113,116-122]. The former approach typically involves using recurrent neural networks to learn spatiotemporal consistency between adjacent frames in an image sequence. The latter approach has broader applications, leveraging the powerful feature extraction capabilities of convolutional neural networks to learn spatiotemporal features in the images.

The major challenge of designing supervised DL reconstruction techniques for dynamic MRI also arises from constructing a fully sampled training database due to the instability (e.g., long breath-hold in segmented cardiac cine imaging), or infeasibility (e.g., real-time cardiac cine and dynamic contrast enhanced imaging). Another challenge lies in the mismatch of temporal resolutions and physiological motion states between the training and test data. For instance, this mismatch may involve training the network using fully sampled segmented cine during breath-hold, while testing with highly undersampled cine without breath-hold.

### 3.9. Quantitative Imaging

Quantitative MRI (qMRI) aims to quantify a single or multiple physical parameters of tissues such as relaxation times (T1 recovery and T2/T2* decay), proton density (PD), diffusion, susceptibility, or magnetization transfer, etc. Based on the principle of "differential weighting" [123], qMRI firstly acquiring multiple contrast weighted images by varying sequence parameters, and subsequently fitting them to the specific physical models to estimate the qMRI parameter values. The challenge of qMRI lies in the long scan time due to acquiring multiple contrast weighted images in a single scan, thus highly demands for acceleration. qMRI from undersampled k-space data can be accomplished in two steps: first image reconstruction and then parameter estimation, or in a single step by physical model-based reconstruction, as illustrated in Eq.7, where $\mathcal{M}$ and $v$ denote the physical model and the unknown physical parameters respectively. Model-based reconstruction provides the advantage of dimensionality reduction of unknows to achieve higher acceleration, albeit at the cost of increased computational complexity and burden due to most physical models of qMRI are nonlinear. Based on the quantitative tissue maps provided by qMRI techniques, qualitative contrast-weighted image can be synthesized by applying tissue maps to a specific physical model without increasing acquisition time, leading to physics driven contrast synthesis.

$$\hat{v} = \min_{v} \|KFC\mathcal{M}(v) - m\|_2^2 + \sum_i \lambda_i \mathcal{R}_i(v) \tag{7}$$

Deep learning has also been investigated to reconstruct quantitative MRI mappings [124-138]. Developing supervised DL reconstruction method for qMRI is more challenging than for conventional MRI since fully sampling multiple contrast weighted images would lead to infeasible scan time [138]. However, the physical model of qMRI and its flexible representations such as analytical equations, dictionary, linear subspace, or even the original Bloch equation provides the opportunities for designing self-supervised or unsupervised DL reconstruction methods. There has been a review paper pointing out the typical four ways to integrate deep learning with the quantitative MRI. More information can be referred to that paper [139].

## 4. Deep Learning Paradigm Shift and Applications for MRI Reconstruction

In the last few years, many DL methods have been introduced for MRI reconstruction, which has seen a paradigm shift from supervised, to semi-supervised and then unsupervised methods, to relieve the reliance on the dataset, to utilize more MRI domain knowledge to empower the deep learning model.

### 4.1. Supervised DL for fast MRI

In MRI reconstruction, supervised learning-based approaches are the most used methods, which require many training sample pairs. **Figure 3** illustrates the general procedure of these methods. In supervised learning, the undersampled k-space data $m_i$ and its corresponding full k-space data reconstructed image $v_i$ are known and are fed into a DL model for training.

Supervised learning-based methods have two main categories, model unrolling-based methods and data-driven methods. The unrolling methods learn optimal, robust parameters, or other complex prior knowledges from big datasets to achieve the best performances for iterative MRI reconstruction models [1,4,140–153]. The data-driven methods trained the model offline on large datasets to extract valuable knowledge such as the regularizers, mapping and denoising, and then used these learned knowledge for online inferring to reconstruct high-quality images from undersampled k-space data [2,3,50,119,154–161]. Different network architectures have been investigated [85,162–185] and uncertainty has also been estimated and introduced to improve the reconstruction performance [186–188]. However, since supervised learning has been well surveyed in several exiting reviews and this journal has word limits requirements, we have put supervised learning in the supporting information. More details can be referred to the review [6–9] or the supporting document.

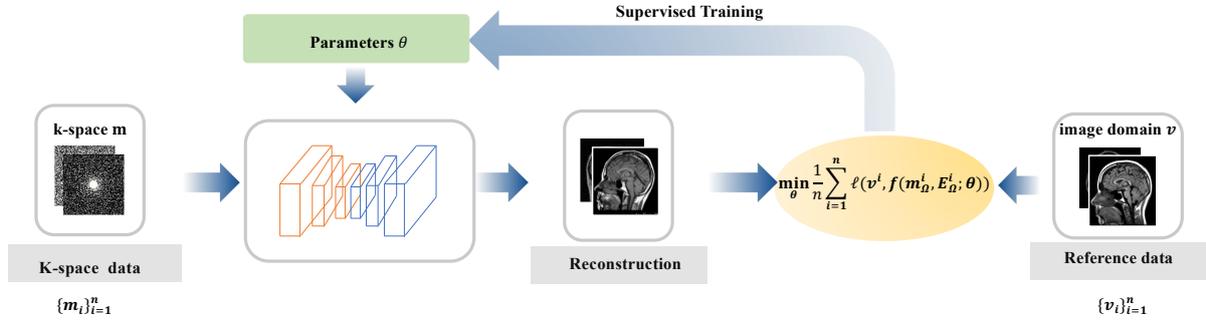

Figure 3

## 4.2. Semi-Supervised DL for fast MRI

Semi-supervised learning is a subfield of machine learning that specializes in using reference and unlabelled or corrupted data to achieve specific learning tasks. The general procedure is illustrated in **Figure 4**. It comprises a supervised learning and an unsupervised learning part, with a limited amount of data for the supervised learning part, while the majority of data is used for the unsupervised learning. The corresponding objective functions are as follows:

$$\min_{\theta} \frac{1}{n}\sum_{i=1}^{n} \ell_1\left(v^i, f(m_\Omega^i, E_\Omega^i; \theta)\right) \tag{8}$$

$$\min_{\theta} \frac{1}{p}\sum_{j=1}^{p} \ell_2\left(\hat{v}^j, f(m_\Omega^j, E_\Omega^j; \theta)\right) \tag{9}$$

where $n$ and $p$ represent the numbers of samples used for supervised and unsupervised training, respectively, with $p$ typically being much larger than $n$. $\hat{v}$ represents pseudo-labels generated by the model trained through supervised learning. Semi-supervised learning-based methods have been explored to solve the lack of references issues.

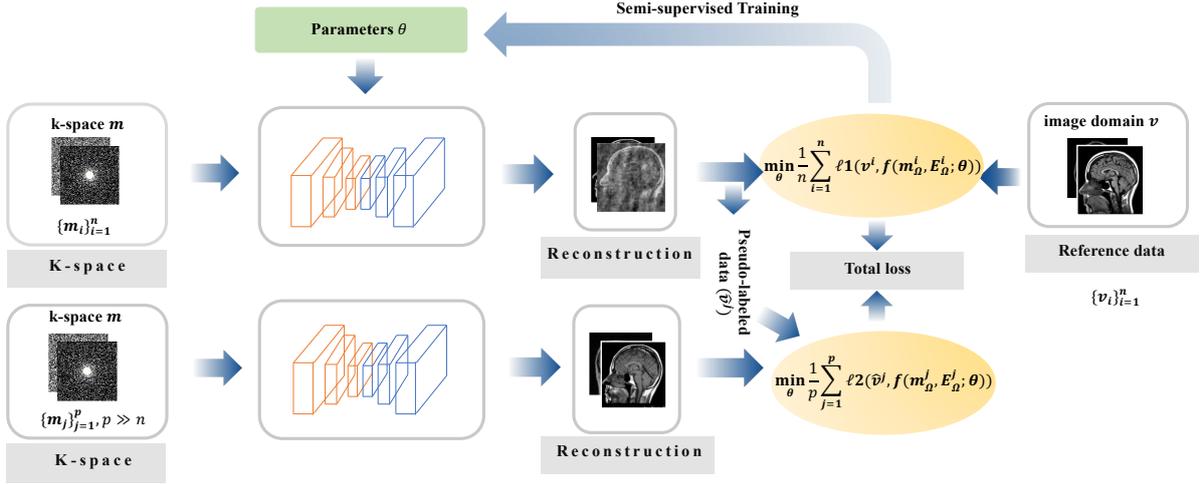

Figure 4

Semi-supervised learning can assist MR image reconstruction from highly accelerated data mainly through three strategies: noise suppression [189], interpolation and filtering[190], and contrast synthesis [191]. For instance, in MRI denoising, the Noise2Recon proposed by Desai et al. [189] is a label-efficient approach that performs joint MRI reconstruction and denoising. It exploits the smoothing assumption of semi-supervised learning. Specifically, the data samples without reference labels are augmented with masked additive noise. The prediction model is robust to noisy out-of-distribution acquisition. VORTEX [192], also belongs to this line of work, which used physical model-driven data augmentation in a semi-supervised consistent training framework to build label-efficient networks that can achieve good reconstruction quality. In super-resolution MRI reconstruction, Mardani et al. [190] introduced a super-resolution MRI reconstruction approach trained in a semi-supervised manner using generative models.

MR contrast synthesis is another popular area where semi-supervised learning is widely used [193-195]. In the clinical practice, only a few contrasts such as T1 and T2 weighted images are acquired in a limited scan time, whereas contrast enhanced T1, one of the most important MRI contrasts for the diagnosis and analysis of tumors, especially gliomas, requires consideration of the additional costs and the risk of allergy to contrast agents. Synthesizing harder-to-obtain image contrasts from other common contrasts becomes a solution. For instance, A semi-supervised generative model, ssGAN, was proposed in the literature [193] for solving the problem of multi-contrast MRI synthesis. Nguyen et al. [194] proposed an SSA-CGAN model that used the adversarial loss to learn from unpaired data points, the cyclic loss to force consistent reconstruction mapping, and another adversarial loss to exploit paired data points, solving the problem of lacking paired data. These methods primarily rely on the generative adversarial networks (GAN) model as the foundational architecture. Leveraging the powerful generative capabilities of GAN models, these methods have achieved excellent reconstruction results. Overall, semi-supervised learning provides an interesting option for assisting MRI reconstruction although there are relatively fewer studies conducted.

**4.3. Unsupervised DL for fast MRI**

The above sections on supervised and semi-supervised learning-based MRI reconstruction methods assume that there are ground-truth MR images to train the model and learn the mapping between the input data and the ground truth. However, it is very expensive or even impossible to collect high-quality fully sampled datasets. Under these circumstances, it is worth investigating how to optimize the deep learning models without the fully sampled datasets. Therefore, unsupervised learning-based methods have been employed to tackle this issue. The objective is to minimize the difference between reconstructed images and the original input images at the undersampled k-space locations. In this review, we classify unsupervised learning approaches for fast MRI into two main categories: self-supervised methods and generative model-based approaches.

Self-supervised learning methods mainly construct scanned undersampled MRI data from the re-undersampled data to achieve network training without reference data, as illustrated in **Figure 5**. The corresponding loss function is represented as:

$$\min_{\theta} \frac{1}{n}\sum_{i=1}^{n} \ell\left(m_{\Lambda}^{i}, E_{\Lambda}^{i}(f(m_{\Psi}^{i}, E_{\Psi}^{i}; \theta))\right) \tag{10}$$

where the subsampled data indices $\Omega$ is divided into two sets $\Psi$ and $\Lambda$. $\Psi$ is used to for reconstruction task, and $\Lambda$ is utilized for computing the loss.

Several studies have investigated self-supervised learning for MRI reconstruction [197-212]. For instance, Yaman et al. [197] introduced a self-supervised approach (SSDU) in which the acquired undersampled data indices are divided into a set of k-space positions used in the network's DC layer during training, and a set of k-space positions used within the loss function. This is a classic work in self-supervised MRI reconstruction, offering valuable insights for subsequent self-supervised learning methods. Although this approach achieved comparable performance to supervised learning-based methods, the information fed to the network for learning was reduced due to the division of the undersampled data. This may result in a poor reconstruction result at high acceleration rates. Thereby, the authors [200] further introduced a multi-mask approach to improve the reconstruction quality at higher acceleration rates. Hu et al. [198] proposed an unsupervised learning framework in which the incomplete k-space data is re-undersampled to construct self-supervision. A difference loss between the parallel networks is used to constrain the search space. The reconstruction results for their method achieved better performance than other methods when the fully sampled dataset is absent. Its performance is even close to the fully supervised deep learning framework. Furthermore, some works have incorporated Bayesian theory into their solution of inverse problems. Pang et al. [208] introduced a self-supervised deep learning recovery approach, which was implemented upon Bayesian deep networks and predicted the uncertainty of the target image by training networks with random weights. Aggarwal et al. [209] introduced ENSURE, an ensemble stein's unbiased risk estimate framework, for unsupervised training of MRI reconstruction models. These methods provide a certain guarantee for the reliability of reconstructed images.

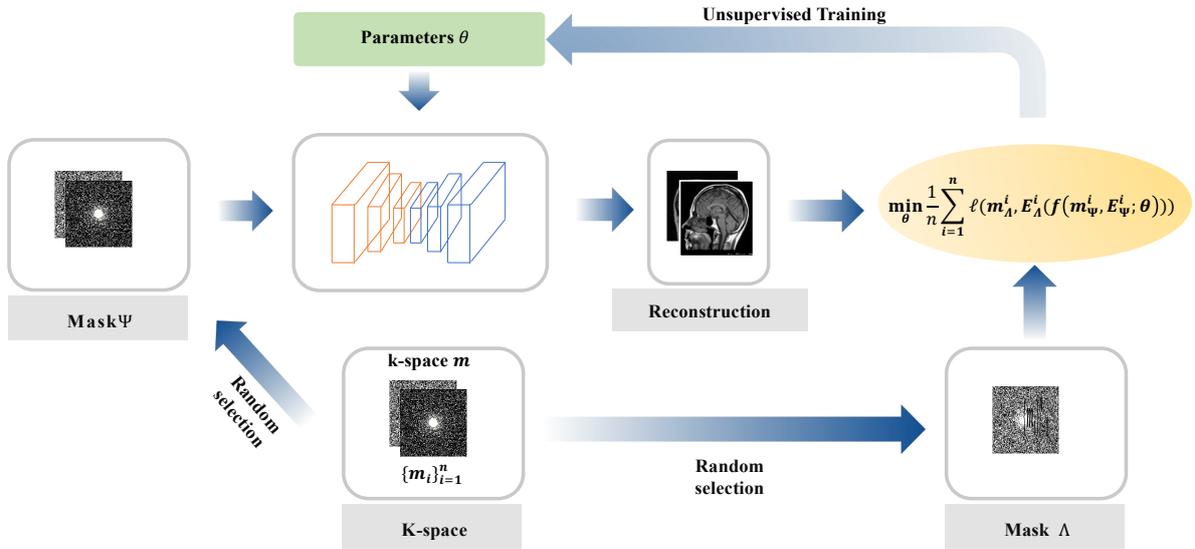

Figure 5

Another widely used unsupervised approach is building generative-based models, which are mainly based on generative adversarial networks (GANs) that take some random distributions or original data as input to generate the corresponding target data [213-225,229]. For example, Xia et al. [213] proposed a robust adversarial learning super-resolution algorithm based on conditional GANs for high-resolution and isotropic cardiac magnetic resonance cine imaging. The algorithm incorporates optical flow components to generate auxiliary images to guide image synthesis and is able to synthesize cardiac MR images that are three-dimensionally isotropic, and anatomically plausible. The problem of model instability was effectively addressed by Lei et al. [214] by performing unpaired adversarial training with the help of Wasserstein GAN (WGAN) and evaluating the quality of the reconstructed images according to the Wasserstein distance. Chung et al. [217] proposed a CycleGAN-based two-stage deep learning method for accelerated 3D TOF MRA, which consisted of a multi-coil reconstruction network in the coronal plane and a multi-planar refinement network in the axial plane.

Additionally, autoencoders have also been used in highly accelerated diffusion MRI [221] and score-based generative models [222]. Tezcan et al. [223] proposed to use variational autoencoders to learn the distributions of fully sampled MR images and extract deep density priors for reconstruction. Score-based diffusion models have also been employed to fit the data distributions step-by-step for accelerated MRI [222]. With the powerful capabilities of generative models, these methods have achieved excellent results. Recently popular diffusion models have also opened up promising directions in this field [96].

In addition to the aforementioned methods, there are other means such as manifold learning/deep subspace learning to address the problem of network training without ground truth images [226-228]. Zou et al. [227] introduced a variational manifold learning method for accelerated dynamic MRI. They also proposed an unsupervised motion-compensated reconstruction scheme (MoCo-StoRM) for pulmonary MRI based on manifold learning [228]. With the powerful modeling capabilities of manifold learning, these methods have achieved excellent performance. Additionally, leveraging prior knowledge from MR data, some methods based on image priors have also been employed in MRI [196,230-237]. For example, Zhao et al. [230] proposed a reference-driven strategy, which was based on the deep image prior (DIP) framework, that can be utilized for MRI reconstruction without requiring any pre-training or training dataset. Similarly, based on DIP, Yoo et al. [232] proposed an unsupervised deep learning method for dynamic MRI. Motivated by DIP, Ahmed et al. [234] designed a bilinear model to decompose dynamic data into spatial and temporal factors for accelerated dynamic imaging. These methods can better preserve the structural information of images and exhibit a certain resistance to noise. In addition, compared to traditional methods, DIP-based approaches can effectively integrate prior knowledge from MR data, enhancing the accuracy of image features and the quality of reconstruction.

The unsupervised training techniques for fast MR imaging have seen great progresses in recent years and their performances are even competitive against supervised-trained deep learning methods.

**4.4. MR Vendors' choices for DL Reconstruction**

The powerful performance of DL techniques demonstrated in CV and image processing, coupled with the surge in academic publications introducing new DL reconstruction methods, have prompted MR vendors to pursue DL-powered solutions for MR image reconstruction.

GE's "AIR Recon DL" product introduces two DL based image processing functions: Gibbs ringing suppression and noise removal to achieve better image quality than the conventional approaches. Network is trained in a supervised manner using pairs of images representing near-perfect images and low-quality images. The complex-valued k-space raw data is utilized to enforce the consistency with the acquired data. The training database spans a broad range of image contents (contrasts, organs, etc.) and is expanded by image augmentations to enable generalizability and robustness across all anatomies. Their network also discards bias terms and employs ReLU activations to make the network to be suitable for MR images of significantly different scales and operate seamlessly with existing acceleration strategies [238].

Philips provides a supervised DL reconstruction technology named as "SmartSpeed". This approach combines iterative algorithm unrolling based Adaptive-CS-Net with domain specific knowledge terms, such as SENSE based data consistency, phase behavior, and background mask. SmartSpeed also utilizes the strong correlation between neighboring slices and a weighted sum of Multiscale-SSIM and L1 as the loss function to improve the images reconstruction fidelity. Adaptive-CS-Net trained on data from the first FastMRI challenge achieved the best scores in the multi-coil 8x accelerated track. Philips separately provides multiple SmartSpeed packages for different organs including brain, spine, MSK, cardiac, and body [239].

Siemens' DL image reconstruction solution named as "Deep Resolve" provides four tools including Boost, Gain, Sharp, and Swift Brain. Deep Resolve Boost applies a deep neural network multiple times in an iterative process of image reconstruction to generate the final output with significantly reduced noise from undersampled raw data. It is claimed that Deep Resolve Boost can be adapted to body regions from head to toe and provided three denoising strengths to the user. Deep Resolve Gain performs denoising with the help of a noise map generated from the raw data to account for the local variation of noise in the image induced by coil geometry and acceleration techniques. Deep Resolve Sharp aims to generates a sharp output from a low-resolution input using a super resolution neural network. The Deep Resolve Swift Brain combines EPI acquisition with a DL reconstruction to achieve ultrafast multi-contrast neuro exam in a single scan of 2 minutes [240].

UIH provides DL powered "ACS" (AI assisted Compressed Sensing) and "DeepRecon" technologies for image reconstruction from undersampled data and for image post-processing tasks respectively. ACS is built upon the conventional iterative CS reconstruction algorithm and unifies the reconstruction models of PF and PI as the data consistency term to improve its interpretability and robustness. DeepRecon utilizes 2D or 3D convolution neural network for 2D/3D image denoising, filtering, and super-resolution to surpass the conventional approaches [241].

Most vendors provide supervised DL solutions tailored for various image processing tasks such as denoising, filtering, and interpolation. These solutions uniformly leverage k-space raw data to enforce consistency with the acquired data. Most DL powered image reconstruction solutions provided by MR vendors are supervised and built upon the conventional CSPI model and iterative algorithms to provide interpretability, and to fully utilize the domain specific experience and knowledge.

In the Sections 3 and 4, we explored the deep learning methods MRI from different perspectives. The Section 3 primarily focuses on the unique properties of MR images, while the Section 4 mainly introduces the paradigm shift of deep learning approaches for MRI reconstruction and the selection of vendors. The unique properties of MR images distinguish them from other types of data. By incorporating the unique domain knowledge (unique properties) of MR images, we can facilitate the transformation of deep learning paradigms used for MRI reconstruction.

## 5. Challenges and Open Questions of DL MR Reconstruction

### 5.1. The Incorporation of Physical Knowledge

Labeled large-scale datasets are crucial for the training of deep learning models. In the field of natural images, there are millions of labeled large-scale datasets such as ImageNet[I], MS-COCO[II], and Open images[III]. These datasets have contributed to the rapid development of the computer vision field. In the field of MRI, due to the physical limitations of imaging devices, acquiring fully sampled MRI data requires a long scanning time, which may impact patient comfort. Additionally, in the acquisition of dynamic imaging data, such as cardiac or abdominal imaging, artifacts may be present in the acquired data due to specific physiological motion. Whereas the commonly used fastMRI[IV] dataset contains only thousands of samples. Therefore, further research is needed to explore methods utilizing few fully sampled data (semi-supervised learning) or without fully sampled data (unsupervised learning). Furthermore, integrating the special properties of MRI into imaging methods, such as complex-value property, dual-space property, multi-contrast property, multi-channel property, and dynamic property, can further reduce dependence on fully sampled data while improving imaging quality. Consequently, further research on how to implement a unified framework that incorporates all these properties in an unsupervised manner is encouraged.

### 5.2. Challenge of Performance Evaluation

---

[I] https://www.image-net.org/
[II] https://cocodataset.org/#home
[III] https://storage.googleapis.com/openimages/web/index.html
[IV] https://fastmri.med.nyu.edu

Currently, the evaluation of DL-based MR reconstruction methods heavily relies on several quantitative measures such as normalized mean square error (NMSE), peak signal to noise ratio (PSNR), structural similarity measurement (SSIM) etc., which cannot ensure the integrity of the diagnosis. Hence, proper clinical validation studies are needed before any application of these methods in real scenarios. There are studies evaluating the reconstructed images from a clinical perspective. For example, Park et al. [242] evaluated the image quality and diagnostic performance accelerated MRI with deep learning-based reconstruction, concluding that deep learning could reduce MRI acquisition time without compromising the image quality. Zochowski et al. [243] evaluated deep learning-based MRI reconstruction for clinical applications when employed for the reconstruction of peripheral nerves. Foreman et al. [52] evaluated and validated the effectiveness of deep learning-based reconstruction models for accelerated ankle imaging. Chan et al. [244] has demonstrated that good quantitative metrics do not necessarily translate into good resolution in reconstructed images. Dziadosz et al. [245] has shown a similar finding – machine learning methods may achieve high quantitative metrics, but those results may be biased. Blau et al. [246] thinks that there exists a trade-off between quantitative distortion measures and perceptual quality. All these provide interesting insights in the evaluation of algorithms.

**5.3. Stability**

The stability of deep learning models is also worth considering. The literature [247] points out that there are severe unstable issues with deep learning MR reconstruction methods. First, tiny perturbations in either spatial or frequential domain may result in severe artefacts in the MR image reconstruction. Second, there is a great variety in the instabilities with respect to structural changes ranging from complete removal of details to more subtle distortions and blurring of the features. The deep learning model doesn't show strong robustness to different sampling patterns, and they have to be retrained on very specific subsampling pattern, subsampling ratio and dimensions used. Last but not least, the literature [247] also mentions a counterintuitive type of instability, the network can yield poorer performances when the samples are more. Therefore, vast efforts are in need to address these instability issues and thus avoid incorrect medical diagnosis if deep learning (especially supervised learning methods) are applied in medical imaging. Additionally, in recent years, there have been theoretical attempts to explain the instability observed in deep learning reconstruction methods, providing a certain research foundation for subsequent researchers [248,249].

**5.4. Implicit Data Crimes**

Another important open issue for deep learning-based MRI reconstruction methods is when developments are performed with retrospective undersampling (simulated undersampling is applied on fully sampled data to get the accelerated acquisition), i.e., the subtle data crime. This can cause problems when moving from retrospectively accelerated acquisitions to prospective acquisitions on real MR imaging systems, leading to biased, overly optimistic results [30]. Moreover, MRI data contains special properties, but only a few are considered in most DL methods for MRI reconstruction. Ignoring such properties can be considered as improper use of validation data as it does not reflect the accurate signals received by the imaging devices, thus can easily lead to incorrect conclusions [250]. For example, MRI data are complex-valued (have both magnitude and phase information). If only the magnitude value is exploited, the accurate signals received by the imaging devices are not used. The performance can be biased. Furthermore, bias could manifest from the choice of technique selected for computing fully sampled images. In parallel MRI, different coil combination techniques could lead to different results [251,252]. Therefore, further attention is required from researchers to carefully consider data usage and experimental design.

## 5.5. No Guarantee of Global Optimality

Although deep learning has demonstrated exceptional performance in MR image reconstruction, surpassing traditional CS-MRI methods and solving numerous MRI reconstruction issues, there remain several challenges yet to be addressed [253]. These challenges extend beyond those previously mentioned. Specifically, there is an issue of confusion regarding optimal solutions, making it difficult to establish a universal neural network for deep learning MR image reconstruction. The ability to obtain a universal model architecture with optimal hyperparameters for a specific problem is still unknown, as there are numerous possible configurations and a lack of theoretical guidelines. It would be valuable to develop a universal deep learning-based MRI system that can handle the entire reconstruction process [254], including data acquisition, image reconstruction, image analysis, multi-tasking models, and deployment. This development would pave the way for faster, more robust, reliable, stable, and accurate MR imaging. Achieving this would require collaborative studies between different vendors, academia, and clinical teams to establish milestones. Determining the optimal network architecture, learning paradigm, and amount of training data required, among other factors, would be crucial in this endeavor. Through this review, we aim to present new avenues for technical development in the field of MRI reconstruction.

## 5.6. Generalizability and Privacy Concerns

One issue with deep learning-based MRI reconstruction models is the generalizability and robustness to different datasets. There is no theoretical guarantee that the model can be generalized to different datasets. To enhance the model's generalizability, researchers tend to use large and comprehensive datasets to train the model, which may in the future cause patient privacy violation issues. A few studies have reported robust performance of the models across different datasets by utilizing federated learning (FL) [255,256] which enables multi-institutional and multi-scanner studies across sites to implement accurate and robust DL techniques without exchanging or revealing data privacy. For example, Elmas et al. [256] introduced FedGIMP, an FL approach based on cross-site learning of a generative MRI prior and prior adaptation to improve patient privacy and model generalizability to multi-site data. Wu et al. [257] recently introduced a novel approach called FedAutoMRI, which combines differentiable architecture search with federated learning. FedAutoMRI automatically searches for the optimal network architecture for MR image reconstruction, enhancing reconstruction accuracy. Additionally, it employs an exponential moving average method to enhance the robustness of the client model in handling data heterogeneity. This work represents the first attempt to use federated neural architecture search for MR image reconstruction. Generally speaking, future studies on FL-based fast MRI will need to tackle issues such as data diversity (data heterogeneity), system architecture (software architecture design in building federated learning systems), and traceability (tracing the real leaker who illegally disclosed the global model to third parties outside FL community). Furthermore, the recently proposed score-based generative or diffusion model brings discoveries in this direction, which improves the generalization of models on unknown measurement processes by capturing the prior distribution of measurement data using a score-based generative or diffusion model. Further studies along these directions are welcome to address generalizability issues in a self-supervised or unsupervised manner.

## 6. Conclusion and Outlook

In this review, we have explored the application of deep learning in MRI reconstruction, which has demonstrated promising results. We have emphasized the importance of domain-specific MR knowledge and potential DL-MRI reconstruction strategies. The DL-MRI methods have experienced a paradigm shift from supervised learning to semi-supervised learning, and finally, to unsupervised learning, depending on the availability of fully sampled reference data. Although these deep learning-based approaches offer powerful learning capabilities and enable real-time imaging, they lack the strong guidance typically found in traditional iterative algorithms. We anticipate that further integration of traditional CS-MRI methods and deep learning techniques will not only guide the development of more generalizable and interpretable deep networks but also, more importantly, yield improved results in real-world scenarios. Additionally, leveraging the information complementarity between multi-contrast images and the redundancy of MRI across parallel imaging channels or higher imaging dimensions could serve as priors to enhance reconstruction performance further. By pursuing these objectives and fostering collaboration among mathematicians, physicians, patients, and imaging scientists, we set milestones for faster, more robust, and more accurate MR imaging.

One notable area of progress is the emergence of federated learning methods. These approaches facilitate collaborative model training across multiple institutions without sharing sensitive patient data. An example of this is the self-supervised federated learning method proposed by Zou et al. [258]. Another area of interest is memory-efficient methods for MRI reconstruction. Wang et al. [259] have introduced a memory-efficient framework specifically designed for high-dimensional MRI reconstruction. Pramanik et al. [260] explores the effectiveness of the monotone operator learning (MOL) framework in parallel MRI acceleration. The MOL algorithm combines a monotone CNN with a conjugate gradient algorithm to ensure data consistency. In addition, leveraging prior information from multi-contrast imaging has been shown to enhance the performance of score-based generative models. Lastly, to prevent suboptimal solutions in diffusion models applied to MRI reconstruction, Mirza et al. [261] have proposed a Fourier-constrained diffusion bridge. This innovative approach ensures that the diffusion models yield optimal results.

These recent developments in MRI reconstruction techniques hold great promise for improving the accuracy, efficiency, and scalability of image reconstruction processes. They have the potential to further advance the medical imaging area.

**Acknowledgements**

This research was partly supported by the National Natural Science Foundation of China (62222118, U22A2040), Guangdong Provincial Key Laboratory of Artificial Intelligence in Medical Image Analysis and Application (2022B1212010011), and Shenzhen Science and Technology Program (RCYX20210706092104034, JCYJ20220531100213029).

**Author contributions**

Shanshan Wang contributed to the design and conceptualization of the overall framework of the review, summarized the special MR properties, and outlined the manuscript; Sen Jia contributed to write the MR imaging physics session and provided the insights about different methods' connections. Ruoyou Wu, Alou Diakite, and Cheng Li contributed to the literature search, organization and analysis. Qiegen Liu and Leslie Ying revised the paper and provided literature review for deep learning methods.

**References**

1. Hammernik K, Klatzer T, Kobler E, et al. Learning a variational network for reconstruction of accelerated MRI data. *Magn Reson Med*. 2018;79(6):3055-3071.
2. Wang S, Su Z, Ying L, et al. Accelerating magnetic resonance imaging via deep learning. In: *2016 IEEE 13th International Symposium on Biomedical Imaging (ISBI)*. IEEE; 2016:514-517.
3. Han Y, Yoo J, Kim HH, Shin HJ, Sung K, Ye JC. Deep learning with domain adaptation for accelerated projection-reconstruction MR. *Magn Reson Med*. 2018;80(3):1189-1205.
4. Aggarwal HK, Mani MP, Jacob M. MoDL: Model-based deep learning architecture for inverse problems. *IEEE Trans Med Imaging*. 2018;38(2):394-405.
5. Yang Y, Sun J, Li H, Xu Z. ADMM-CSNet: A deep learning approach for image compressive sensing. *IEEE Trans Pattern Anal Mach Intell*. 2018;42(3):521-538.
6. Wang S, Xiao T, Liu Q, Zheng H. Deep learning for fast MR imaging: a review for learning reconstruction from incomplete k-space data. *Biomed Signal Process Control*. 2021;68:102579.
7. Knoll F, Hammernik K, Zhang C, et al. Deep-learning methods for parallel magnetic resonance imaging reconstruction: A survey of the current approaches, trends, and issues. *IEEE Signal Process Mag*. 2020;37(1):128-140.
8. Liang D, Cheng J, Ke Z, Ying L. Deep magnetic resonance image reconstruction: Inverse problems meet neural networks. *IEEE Signal Process Mag*. 2020;37(1):141-151.
9. Chen Y, Schönlieb CB, Lio P, et al. AI-based reconstruction for fast MRI—A systematic review and meta-analysis. *Proc IEEE*. 2022;110(2):224-245.
10. Lam F, Peng X, Liang ZP. High-Dimensional MR Spatiospectral Imaging by Integrating Physics-Based Modeling and Data-Driven Machine Learning: Current progress and future directions. *IEEE Signal Process Mag*. 2023;40(2):101-115.
11. Iqbal Z, Nguyen D, Hangel G, Motyka S, Bogner W, Jiang S. Super-resolution 1H magnetic resonance spectroscopic imaging utilizing deep learning. *Front Oncol*. 2019;9:1010.
12. Mani M, Magnotta VA, Jacob M. qModeL: A plug-and-play model-based reconstruction for highly accelerated multi-shot diffusion MRI using learned priors. *Magn Reson Med*. 2021;86(2):835-851.
13. LeCun Y, Bengio Y, Hinton G. Deep learning. *nature*, 2015, 521(7553): 436-444.
14. Bishop CM, Nasrabadi NM. *Pattern Recognition and Machine Learning*. Vol 4. Springer; 2006.
15. Chapelle O, Scholkopf B, Zien A. Semi-supervised learning (chapelle, o. et al., eds.; 2006)[book reviews]. *IEEE Trans Neural Networks*. 2009;20(3):542.
16. Goodfellow I, Bengio Y, Courville A. *Deep Learning*.; 2016.
17. Blumenthal M, Luo G, Schilling M, Holme HCM, Uecker M. Deep, deep learning with BART. *Magn Reson Med*. 2023;89(2):678-693.
18. Huang Z, Lin L, Cheng P, Pan K, Tang X. DS 3-Net: Difficulty-Perceived Common-to-T1ce Semi-supervised Multimodal MRI Synthesis Network. In: *Medical Image Computing and Computer Assisted Intervention–MICCAI 2022: 25th International Conference, Singapore, September 18–22, 2022,*

*Proceedings, Part VI*. Springer; 2022:571-581.

19. Bloch F. Nuclear induction. *Phys Rev*. 1946;70(7-8):460.

20. Akcakaya M, Doneva MI, Prieto C. *Magnetic Resonance Image Reconstruction: Theory, Methods, and Applications*. Elsevier Science & Technology; 2022.

21. Eo T, Jun Y, Kim T, Jang J, Lee H, Hwang D. KIKI-net: cross-domain convolutional neural networks for reconstructing undersampled magnetic resonance images. *Magn Reson Med*. 2018;80(5):2188-2201.

22. Souza R, Lebel RM, Frayne R. A hybrid, dual domain, cascade of convolutional neural networks for magnetic resonance image reconstruction. In: *International Conference on Medical Imaging with Deep Learning*. PMLR; 2019:437-446.

23. Wang S, Ke Z, Cheng H, et al. DIMENSION: dynamic MR imaging with both k-space and spatial prior knowledge obtained via multi-supervised network training. *NMR Biomed*. 2022;35(4):e4131.

24. Zhou B, Zhou SK. DuDoRNet: learning a dual-domain recurrent network for fast MRI reconstruction with deep T1 prior. In: *Proceedings of the IEEE/CVF Conference on Computer Vision and Pattern Recognition*. ; 2020:4273-4282.

25. Souza R, Bento M, Nogovitsyn N, et al. Dual-domain cascade of U-nets for multi-channel magnetic resonance image reconstruction. *Magn Reson Imaging*. 2020;71:140-153.

26. Ran M, Xia W, Huang Y, et al. Md-recon-net: A parallel dual-domain convolutional neural network for compressed sensing mri. *IEEE Trans Radiat Plasma Med Sci*. 2020;5(1):120-135.

27. Liu X, Pang Y, Jin R, Liu Y, Wang Z. Dual-domain reconstruction network with V-Net and K-Net for fast MRI. *Magn Reson Med*. 2022;88(6):2694-2708.

28. Liu Y, Pang Y, Liu X, Liu Y, Nie J. DIIK-Net: A full-resolution cross-domain deep interaction convolutional neural network for MR image reconstruction. *Neurocomputing*. 2023;517:213-222.

29. Zhao X, Yang T, Li B, Zhang X. SwinGAN: A dual-domain Swin Transformer-based generative adversarial network for MRI reconstruction. *Comput Biol Med*. 2023;153:106513.

30. Shimron E, Tamir JI, Wang K, Lustig M. Implicit data crimes: Machine learning bias arising from misuse of public data. *Proc Natl Acad Sci*. 2022;119(13):e2117203119.

31. Han Y, Sunwoo L, Ye JC. K-space deep learning for accelerated MRI. *IEEE Trans Med Imaging*. 2019;39(2):377-386.

32. Wymer DT, Patel KP, Burke III WF, Bhatia VK. Phase-contrast MRI: physics, techniques, and clinical applications. *Radiographics*. 2020;40(1):122-140.

33. Lee D, Yoo J, Tak S, Ye JC. Deep residual learning for accelerated MRI using magnitude and phase networks. *IEEE Trans Biomed Eng*. 2018;65(9):1985-1995.

34. Feng CM, Yang Z, Fu H, Xu Y, Yang J, Shao L. DONet: dual-octave network for fast MR image reconstruction. *IEEE Trans Neural Networks Learn Syst*. 2021.

35. Xiao L, Liu Y, Yi Z, et al. Partial Fourier reconstruction of complex MR images using complex-valued convolutional neural networks. *Magn Reson Med*. 2022;87(2):999-1014.

36. Cole E, Cheng J, Pauly J, Vasanawala S. Analysis of deep complex-valued convolutional neural networks for MRI reconstruction and phase-focused applications. *Magn Reson Med*. 2021;86(2):1093-1109.

37. Wang S, Cheng H, Ying L, et al. DeepcomplexMRI: Exploiting deep residual network for fast parallel MR imaging with complex convolution. *Magn Reson Imaging*. 2020;68:136-147.

38. El-Rewaidy H, Neisius U, Mancio J, et al. Deep complex convolutional network for fast reconstruction of 3D late gadolinium enhancement cardiac MRI. *NMR Biomed*. 2020;33(7):e4312.
39. Dedmari MA, Conjeti S, Estrada S, Ehses P, Stöcker T, Reuter M. Complex fully convolutional neural networks for MR image reconstruction. In: *International Workshop on Machine Learning for Medical Image Reconstruction*. Springer; 2018:30-38.
40. Pruessmann KP, Weiger M, Scheidegger MB, Boesiger P. SENSE: sensitivity encoding for fast MRI. *Magn Reson Med An Off J Int Soc Magn Reson Med*. 1999;42(5):952-962.
41. Griswold MA, Jakob PM, Heidemann RM, et al. Generalized autocalibrating partially parallel acquisitions (GRAPPA). *Magn Reson Med An Off J Int Soc Magn Reson Med*. 2002;47(6):1202-1210.
42. Lustig M, Pauly JM. SPIRiT: iterative self-consistent parallel imaging reconstruction from arbitrary k-space. *Magn Reson Med*. 2010;64(2):457-471.
43. Zhang J, Liu C, Moseley ME. Parallel reconstruction using null operations. *Magn Reson Med*. 2011;66(5):1241-1253.
44. Uecker M, Lai P, Murphy MJ, et al. ESPIRiT—an eigenvalue approach to autocalibrating parallel MRI: where SENSE meets GRAPPA. *Magn Reson Med*. 2014;71(3):990-1001.
45. Hu J, Yi Z, Zhao Y, et al. Parallel imaging reconstruction using spatial nulling maps. *Magn Reson Med*. 2023;(90):502-519. doi:10.1002/mrm.29658
46. Shin PJ, Larson PEZ, Ohliger MA, et al. Calibrationless parallel imaging reconstruction based on structured low-rank matrix completion. *Magn Reson Med*. 2014;72(4):959-970.
47. Kim TH, Setsompop K, Haldar JP. LORAKS makes better SENSE: phase-constrained partial fourier SENSE reconstruction without phase calibration. *Magn Reson Med*. 2017;77(3):1021-1035.
48. Jin KH, Ye JC. Annihilating filter-based low-rank Hankel matrix approach for image inpainting. *IEEE Trans Image Process*. 2015;24(11):3498-3511.
49. Pawar K, Egan GF, Chen Z. Domain knowledge augmentation of parallel MR image reconstruction using deep learning. *Comput Med Imaging Graph*. 2021;92:101968.
50. Kwon K, Kim D, Park H. A parallel MR imaging method using multilayer perceptron. *Med Phys*. 2017;44(12):6209-6224.
51. Peng X, Sutton BP, Lam F, Liang Z. DeepSENSE: Learning coil sensitivity functions for SENSE reconstruction using deep learning. *Magn Reson Med*. 2022;87(4):1894-1902.
52. Foreman SC, Neumann J, Han J, et al. Deep learning–based acceleration of Compressed Sense MR imaging of the ankle. *Eur Radiol*. 2022;32(12):8376-8385.
53. Arvinte M, Vishwanath S, Tewfik AH, Tamir JI. Deep J-Sense: Accelerated MRI reconstruction via unrolled alternating optimization. In: *International Conference on Medical Image Computing and Computer-Assisted Intervention*. Springer; 2021:350-360.
54. Lu T, Zhang X, Huang Y, et al. pFISTA-SENSE-ResNet for parallel MRI reconstruction. *J Magn Reson*. 2020;318:106790.
55. Zhang X, Lu H, Guo D, et al. A guaranteed convergence analysis for the projected fast iterative soft-thresholding algorithm in parallel MRI. *Med Image Anal*. 2021;69:101987.
56. Hammernik K, Schlemper J, Qin C, Duan J, Summers RM, Rueckert D. Systematic evaluation of iterative deep neural networks for fast parallel MRI reconstruction with sensitivity-weighted coil combination. *Magn Reson Med*. 2021;86(4):1859-1872.

57. Chen Y, Xiao T, Li C, Liu Q, Wang S. Model-based convolutional de-aliasing network learning for parallel MR imaging. In: *Medical Image Computing and Computer Assisted Intervention–MICCAI 2019: 22nd International Conference, Shenzhen, China, October 13–17, 2019, Proceedings, Part III 22*. Springer; 2019:30-38.

58. Zhou Z, Han F, Ghodrati V, et al. Parallel imaging and convolutional neural network combined fast MR image reconstruction: Applications in low-latency accelerated real-time imaging. *Med Phys*. 2019;46(8):3399-3413.

59. Lv J, Li G, Tong X, et al. Transfer learning enhanced generative adversarial networks for multi-channel MRI reconstruction. *Comput Biol Med*. 2021;134:104504.

60. Jun Y, Shin H, Eo T, Hwang D. Joint deep model-based MR image and coil sensitivity reconstruction network (joint-ICNet) for fast MRI. In: *Proceedings of the IEEE/CVF Conference on Computer Vision and Pattern Recognition*. ; 2021:5270-5279.

61. Dawood P, Breuer F, Stebani J, et al. Iterative training of robust k-space interpolation networks for improved image reconstruction with limited scan specific training samples. *Magn Reson Med*. 2023;89(2):812-827.

62. Chen F, Cheng JY, Taviani V, et al. Data-driven self-calibration and reconstruction for non-cartesian wave-encoded single-shot fast spin echo using deep learning. *J Magn Reson Imaging*. 2020;51(3):841-853.

63. Hosseini SAH, Zhang C, Weingärtner S, et al. Accelerated coronary MRI with sRAKI: A database-free self-consistent neural network k-space reconstruction for arbitrary undersampling. *PLoS One*. 2020;15(2):e0229418.

64. Zhang C, Moeller S, Demirel OB, Uğurbil K, Akçakaya M. Residual RAKI: A hybrid linear and non-linear approach for scan-specific k-space deep learning. *Neuroimage*. 2022;256:119248.

65. Arefeen Y, Beker O, Cho J, Yu H, Adalsteinsson E, Bilgic B. Scan-specific artifact reduction in k-space (SPARK) neural networks synergize with physics-based reconstruction to accelerate MRI. *Magn Reson Med*. 2022;87(2):764-780.

66. Breuer FA, Blaimer M, Mueller MF, et al. Controlled aliasing in volumetric parallel imaging (2D CAIPIRINHA). *Magn Reson Med An Off J Int Soc Magn Reson Med*. 2006;55(3):549-556.

67. Murphy M, Alley M, Demmel J, Keutzer K, Vasanawala S, Lustig M. Fast l1-SPIRiT compressed sensing parallel imaging MRI: scalable parallel implementation and clinically feasible runtime. *IEEE Trans Med Imaging*. 2012;31(6):1250-1262.

68. Levine E, Hargreaves B. On-the-Fly Adaptive ${k}$-Space Sampling for Linear MRI Reconstruction Using Moment-Based Spectral Analysis. *IEEE Trans Med Imaging*. 2017;37(2):557-567.

69. Aggarwal HK, Jacob M. J-MoDL: Joint model-based deep learning for optimized sampling and reconstruction. *IEEE J Sel Top Signal Process*. 2020;14(6):1151-1162.

70. Wright KL, Hamilton JI, Griswold MA, Gulani V, Seiberlich N. Non-Cartesian parallel imaging reconstruction. *J Magn Reson Imaging*. 2014;40(5):1022-1040.

71. Malavé MO, Baron CA, Koundinyan SP, et al. Reconstruction of undersampled 3D non-Cartesian image-based navigators for coronary MRA using an unrolled deep learning model. *Magn Reson Med*. 2020;84(2):800-812.

72. Schlemper J, Salehi SSM, Kundu P, et al. Nonuniform variational network: deep learning for accelerated

nonuniform MR image reconstruction. In: *Medical Image Computing and Computer Assisted Intervention–MICCAI 2019: 22nd International Conference, Shenzhen, China, October 13–17, 2019, Proceedings, Part III 22*. Springer; 2019:57-64.

73. Zhang J, Zhang H, Wang A, et al. Extending LOUPE for K-space Under-sampling Pattern Optimization in Multi-coil MRI. In: *Machine Learning for Medical Image Reconstruction: Third International Workshop, MLMIR 2020, Held in Conjunction with MICCAI 2020, Lima, Peru, October 8, 2020, Proceedings 3*. Springer; 2020:91-101.

74. Ramzi Z, Starck JL, Ciuciu P. Density compensated unrolled networks for non-cartesian MRI reconstruction. In: *2021 IEEE 18th International Symposium on Biomedical Imaging (ISBI)*. IEEE; 2021:1443-1447.

75. Liu F, Samsonov A, Chen L, Kijowski R, Feng L. SANTIS: sampling-augmented neural network with incoherent structure for MR image reconstruction. *Magn Reson Med*. 2019;82(5):1890-1904.

76. Ramzi Z, Chaithya GR, Starck JL, Ciuciu P. NC-PDNet: A density-compensated unrolled network for 2D and 3D non-Cartesian MRI reconstruction. *IEEE Trans Med Imaging*. 2022;41(7):1625-1638.

77. Wang G, Luo T, Nielsen JF, Noll DC, Fessler JA. B-spline parameterized joint optimization of reconstruction and k-space trajectories (bjork) for accelerated 2d mri. *IEEE Trans Med Imaging*. 2022;41(9):2318-2330.

78. Bahadir CD, Wang AQ, Dalca A V, Sabuncu MR. Deep-learning-based optimization of the under-sampling pattern in MRI. *IEEE Trans Comput Imaging*. 2020;6:1139-1152.

79. Radhakrishna CG, Ciuciu P. Jointly learning non-cartesian k-space trajectories and reconstruction networks for 2D and 3D MR imaging through projection. *Bioengineering*. 2023;10(2):158.

80. Weiss T, Senouf O, Vedula S, Michailovich O, Zibulevsky M, Bronstein A. PILOT: Physics-informed learned optimized trajectories for accelerated MRI. *arXiv Prepr arXiv190905773*. 2019.

81. Peng W, Feng L, Zhao G, Liu F. Learning optimal k-space acquisition and reconstruction using physics-informed neural networks. In: *Proceedings of the IEEE/CVF Conference on Computer Vision and Pattern Recognition*. ; 2022:20794-20803.

82. Wang J, Yang Q, Yang Q, Xu L, Cai C, Cai S. Joint optimization of Cartesian sampling patterns and reconstruction for single-contrast and multi-contrast fast magnetic resonance imaging. *Comput Methods Programs Biomed*. 2022;226:107150.

83. Sherry F, Benning M, De los Reyes JC, et al. Learning the sampling pattern for MRI. *IEEE Trans Med Imaging*. 2020;39(12):4310-4321.

84. Zibetti MVW, Knoll F, Regatte RR. Alternating learning approach for variational networks and undersampling pattern in parallel MRI applications. *IEEE Trans Comput Imaging*. 2022;8:449-461.

85. Chatterjee S, Breitkopf M, Sarasaen C, et al. Reconresnet: Regularised residual learning for mr image reconstruction of undersampled cartesian and radial data. *Comput Biol Med*. 2022;143:105321.

86. Huang J, Chen C, Axel L. Fast multi-contrast MRI reconstruction. *Magn Reson Imaging*. 2014; 32(10): 1344-1352.

87. Kim KH, Do W, Park S. Improving resolution of MR images with an adversarial network incorporating images with different contrast. *Med Phys*. 2018;45(7):3120-3131.

88. Sun L, Fan Z, Fu X, Huang Y, Ding X, Paisley J. A deep information sharing network for multi-contrast compressed sensing MRI reconstruction. *IEEE Trans Image Process*. 2019;28(12):6141-6153.

89. Do W, Seo S, Han Y, Ye JC, Choi SH, Park S. Reconstruction of multicontrast MR images through deep learning. *Med Phys*. 2020;47(3):983-997.

90. Liu X, Zhang M, Liu Q, et al. Multi-contrast MR reconstruction with enhanced denoising autoencoder prior learning. In: *2020 IEEE 17th International Symposium on Biomedical Imaging (ISBI)*. IEEE; 2020:1-5.

91. Feng CM, Yan Y, Chen G, et al. Multi-modal transformer for accelerated mr imaging. *IEEE Trans Med Imaging*. 2022.

92. Johnson PM, Tong A, Donthireddy A, et al. Deep learning reconstruction enables highly accelerated biparametric MR imaging of the prostate. *J Magn Reson Imaging*. 2022;56(1):184-195.

93. Liu X, Wang J, Lin S, Crozier S, Liu F. Optimizing multicontrast MRI reconstruction with shareable feature aggregation and selection. *NMR Biomed*. 2021;34(8):e4540.

94. Polak D, Cauley S, Bilgic B, et al. Joint multi-contrast variational network reconstruction (jVN) with application to rapid 2D and 3D imaging. *Magn Reson Med*. 2020;84(3):1456-1469.

95. Liu X, Wang J, Sun H, Chandra SS, Crozier S, Liu F. On the regularization of feature fusion and mapping for fast MR multi-contrast imaging via iterative networks. *Magn Reson Imaging*. 2021;77:159-168.

96. Özbey M, Dalmaz O, Dar S U H, et al. Unsupervised medical image translation with adversarial diffusion models. *IEEE Trans Med Imaging*. 2023; 42(12): 3524-3539.

97. Schlemper J, Caballero J, Hajnal J, Price AN, Rueckert D. A deep cascade of convolutional neural networks for dynamic MR image reconstruction. *IEEE Trans Med Imaging*. 2017;37:491-503.

98. Hauptmann A, Arridge S, Lucka F, Muthurangu V, Steeden JA. Real-time cardiovascular MR with spatio-temporal artifact suppression using deep learning–proof of concept in congenital heart disease. *Magn Reson Med*. 2019;81(2):1143-1156.

99. Qin C, Schlemper J, Caballero J, Price AN, Hajnal J V, Rueckert D. Convolutional recurrent neural networks for dynamic MR image reconstruction. *IEEE Trans Med Imaging*. 2018;38(1):280-290.

100. Qin C, Duan J, Hammernik K, et al. Complementary time-frequency domain networks for dynamic parallel MR image reconstruction. *Magn Reson Med*. 2021;86(6):3274-3291.

101. Seegoolam G, Schlemper J, Qin C, Price A, Hajnal J, Rueckert D. Exploiting motion for deep learning reconstruction of extremely-undersampled dynamic MRI. In: *Medical Image Computing and Computer Assisted Intervention–MICCAI 2019: 22nd International Conference, Shenzhen, China, October 13–17, 2019, Proceedings, Part IV*. Springer; 2019:704-712.

102. Biswas S, Aggarwal HK, Jacob M. Dynamic MRI using model-based deep learning and SToRM priors: MoDL-SToRM. *Magn Reson Med*. 2019;82(1):485-494.

103. Sandino CM, Lai P, Vasanawala SS, Cheng JY. Accelerating cardiac cine MRI using a deep learning-based ESPIRiT reconstruction. *Magn Reson Med*. 2021;85(1):152-167.

104. Küstner T, Fuin N, Hammernik K, et al. CINENet: deep learning-based 3D cardiac CINE MRI reconstruction with multi-coil complex-valued 4D spatio-temporal convolutions. *Sci Rep*. 2020;10(1):13710.

105. Fuin N, Bustin A, Küstner T, et al. A multi-scale variational neural network for accelerating motion-compensated whole-heart 3D coronary MR angiography. *Magn Reson Imaging*. 2020;70:155-167.

106. El-Rewaidy H, Fahmy AS, Pashakhanloo F, et al. Multi-domain convolutional neural network (MD-

CNN) for radial reconstruction of dynamic cardiac MRI. *Magn Reson Med*. 2021;85(3):1195-1208.

107. Kofler A, Dewey M, Schaeffter T, Wald C, Kolbitsch C. Spatio-temporal deep learning-based undersampling artefact reduction for 2D radial cine MRI with limited training data. *IEEE Trans Med Imaging*. 2019;39(3):703-717.

108. Terpstra ML, Maspero M, d'Agata F, et al. Deep learning-based image reconstruction and motion estimation from undersampled radial k-space for real-time MRI-guided radiotherapy. *Phys Med Biol*. 2020;65(15):155015.

109. Freedman JN, Gurney-Champion OJ, Nill S, et al. Rapid 4D-MRI reconstruction using a deep radial convolutional neural network: Dracula. *Radiother Oncol*. 2021;159:209-217.

110. Ke Z, Cui ZX, Huang W, et al. Deep manifold learning for dynamic MR imaging. *IEEE Trans Comput Imaging*. 2021;7:1314-1327.

111. Ke Z, Huang W, Cui ZX, et al. Learned low-rank priors in dynamic MR imaging. *IEEE Trans Med Imaging*. 2021;40(12):3698-3710.

112. Cheng J, Cui ZX, Huang W, et al. Learning data consistency and its application to dynamic MR imaging. *IEEE Trans Med Imaging*. 2021;40(11):3140-3153.

113. Huang Q, Xian Y, Yang D, et al. Dynamic MRI reconstruction with end-to-end motion-guided network. *Med Image Anal*. 2021;68:101901.

114. Jaubert O, Montalt-Tordera J, Knight D, et al. Real-time deep artifact suppression using recurrent U-nets for low-latency cardiac MRI. *Magn Reson Med*. 2021;86(4):1904-1916.

115. Lyu Q, Shan H, Xie Y, et al. Cine cardiac MRI motion artifact reduction using a recurrent neural network. *IEEE Trans Med Imaging*. 2021;40(8):2170-2181.

116. Oscanoa JA, Middione MJ, Syed AB, Sandino CM, Vasanawala SS, Ennis DB. Accelerated two-dimensional phase-contrast for cardiovascular MRI using deep learning-based reconstruction with complex difference estimation. *Magn Reson Med*. 2023;89(1):356-369.

117. Chen Z, Chen Y, Xie Y, Li D, Christodoulou AG. Data-Consistent non-Cartesian deep subspace learning for efficient dynamic MR image reconstruction. In: *2022 IEEE 19th International Symposium on Biomedical Imaging (ISBI)*. IEEE; 2022:1-5.

118. Wang Z, She H, Zhang Y, Du YP. Parallel non-Cartesian spatial-temporal dictionary learning neural networks (stDLNN) for accelerating 4D-MRI. *Med Image Anal*. 2023;84:102701.

119. Kofler A, Haltmeier M, Schaeffter T, Kolbitsch C. An end-to-end-trainable iterative network architecture for accelerated radial multi-coil 2D cine MR image reconstruction. *Med Phys*. 2021;48(5):2412-2425.

120. Sandino CM, Ong F, Iyer SS, Bush A, Vasanawala S. Deep subspace learning for efficient reconstruction of spatiotemporal imaging data. In: *NeurIPS 2021 Workshop on Deep Learning and Inverse Problems*. ; 2021.

121. Qi H, Hajhosseiny R, Cruz G, et al. End-to-end deep learning nonrigid motion-corrected reconstruction for highly accelerated free-breathing coronary MRA. *Magn Reson Med*. 2021;86(4):1983-1996.

122. Djebra Y, Marin T, Han PK, Bloch I, El Fakhri G, Ma C. Manifold learning via linear tangent space alignment (LTSA) for accelerated dynamic MRI with sparse sampling. *IEEE Trans Med Imaging*. 2022;42(1):158-169.

123. Jara H, Sakai O, Farrher E, et al. Primary Multiparametric Quantitative Brain MRI: State-of-the-Art Relaxometric and Proton Density Mapping Techniques. *Radiology*. 2022;305(1):5-18.

124. Yoon J, Gong E, Chatnuntawech I, et al. Quantitative susceptibility mapping using deep neural network: QSMnet. *Neuroimage*. 2018;179:199-206.

125. Cai C, Wang C, Zeng Y, et al. Single-shot T2 mapping using overlapping-echo detachment planar imaging and a deep convolutional neural network. *Magn Reson Med*. 2018;80(5):2202-2214.

126. Liu F, Feng L, Kijowski R. MANTIS: model-augmented neural network with incoherent k-space sampling for efficient MR parameter mapping. *Magn Reson Med*. 2019;82(1):174-188.

127. Zhang J, Wu J, Chen S, et al. Robust single-shot T 2 mapping via multiple overlapping-echo acquisition and deep neural network. *IEEE Trans Med Imaging*. 2019;38(8):1801-1811.

128. Liu F, Kijowski R, Feng L, El Fakhri G. High-performance rapid MR parameter mapping using model-based deep adversarial learning. *Magn Reson Imaging*. 2020;74:152-160.

129. Wu Y, Ma Y, Du J, Xing L. Accelerating quantitative MR imaging with the incorporation of B1 compensation using deep learning. *Magn Reson Imaging*. 2020;72:78-86.

130. Hermann I, Martínez-Heras E, Rieger B, et al. Accelerated white matter lesion analysis based on simultaneous T1 and T2∗ quantification using magnetic resonance fingerprinting and deep learning. *Magn Reson Med*. 2021;86(1):471-486.

131. Zhang C, Karkalousos D, Bazin PL, et al. A unified model for reconstruction and R2* mapping of accelerated 7T data using the quantitative recurrent inference machine. *Neuroimage*. 2022;264:119680.

132. Maier O, Schoormans J, Schloegl M, et al. Rapid T1 quantification from high resolution 3D data with model-based reconstruction. *Magn Reson Med*. 2019;81(3):2072-2089.

133. Feng R, Zhao J, Wang H, et al. MoDL-QSM: Model-based deep learning for quantitative susceptibility mapping. *Neuroimage*. 2021;240:118376.

134. Jun Y, Shin H, Eo T, Kim T, Hwang D. Deep model-based magnetic resonance parameter mapping network (DOPAMINE) for fast T1 mapping using variable flip angle method. *Med Image Anal*. 2021;70:102017.

135. Liu F, Kijowski R, El Fakhri G, Feng L. Magnetic resonance parameter mapping using model-guided self-supervised deep learning. *Magn Reson Med*. 2021;85(6):3211-3226.

136. Gao Y, Cloos M, Liu F, Crozier S, Pike GB, Sun H. Accelerating quantitative susceptibility and R2* mapping using incoherent undersampling and deep neural network reconstruction. *Neuroimage*. 2021;240:118404.

137. Li Y, Wang Y, Qi H, et al. Deep learning–enhanced T1 mapping with spatial-temporal and physical constraint. *Magn Reson Med*. 2021;86(3):1647-1661.

138. Shafieizargar B, Byanju R, Sijbers J, Klein S, den Dekker AJ, Poot DHJ. Systematic review of reconstruction techniques for accelerated quantitative MRI. *Magn Reson Med*. 2023;90(3):1172-1208.

139 Zhu Y, Cheng J, Cui Z X, et al. Physics-Driven Deep Learning Methods for Fast Quantitative Magnetic Resonance Imaging: Performance improvements through integration with deep neural networks. *IEEE Signal Process Mag*. 2023; 40(2): 116-128.

140. Zhang X, Lian Q, Yang Y, Su Y. A deep unrolling network inspired by total variation for compressed sensing MRI. *Digit Signal Process*. 2020;107:102856.

141. Lahiri A, Wang G, Ravishankar S, Fessler JA. Blind primed supervised (blips) learning for mr image reconstruction. *IEEE Trans Med Imaging*. 2021;40(11):3113-3124.

142. Hu D, Zhang Y, Zhu J, Liu Q, Chen Y. TRANS-Net: Transformer-Enhanced Residual-Error AlterNative

Suppression Network for MRI Reconstruction. *IEEE Trans Instrum Meas*. 2022;71:1-13.

143. Fabian Z, Tinaz B, Soltanolkotabi M. Humus-net: Hybrid unrolled multi-scale network architecture for accelerated mri reconstruction. *Adv Neural Inf Process Syst*. 2022;35:25306-25319.

144. Yi Z, Hu J, Zhao Y, et al. Fast and Calibrationless low-rank parallel imaging reconstruction through unrolled deep learning estimation of multi-channel spatial support maps. *IEEE Trans Med Imaging*. 2023.

145. Wang Z, Qian C, Guo D, et al. One-dimensional deep low-rank and sparse network for accelerated MRI. *IEEE Trans Med Imaging*. 2022;42(1):79-90.

146. Knoll F, Hammernik K, Kobler E, Pock T, Recht MP, Sodickson DK. Assessment of the generalization of learned image reconstruction and the potential for transfer learning. *Magn Reson Med*. 2019;81(1):116-128.

147. Chen F, Taviani V, Malkiel I, et al. Variable-density single-shot fast spin-echo MRI with deep learning reconstruction by using variational networks. *Radiology*. 2018;289(2):366-373.

148. Hosseini SAH, Yaman B, Moeller S, Hong M, Akçakaya M. Dense recurrent neural networks for accelerated MRI: History-cognizant unrolling of optimization algorithms. *IEEE J Sel Top Signal Process*. 2020;14(6):1280-1291.

149. Zhang M, Li M, Zhou J, et al. High-dimensional embedding network derived prior for compressive sensing MRI reconstruction. *Med Image Anal*. 2020;64:101717.

150. Cheng J, Wang H, Ying L, Liang D. Model learning: Primal dual networks for fast MR imaging. In: *Medical Image Computing and Computer Assisted Intervention–MICCAI 2019: 22nd International Conference, Shenzhen, China, October 13–17, 2019, Proceedings, Part III 22*. Springer; 2019:21-29.

151. Duan J, Schlemper J, Qin C, et al. VS-Net: Variable splitting network for accelerated parallel MRI reconstruction. In: *Medical Image Computing and Computer Assisted Intervention–MICCAI 2019: 22nd International Conference, Shenzhen, China, October 13–17, 2019, Proceedings, Part IV 22*. Springer; 2019:713-722.

152. Pezzotti N, Yousefi S, Elmahdy MS, et al. An adaptive intelligence algorithm for undersampled knee MRI reconstruction. *IEEE Access*. 2020;8:204825-204838.

153. Liu Y, Liu Q, Zhang M, Yang Q, Wang S, Liang D. IFR-Net: Iterative feature refinement network for compressed sensing MRI. *IEEE Trans Comput Imaging*. 2019;6:434-446.

154. Kofler A, Haltmeier M, Schaeffter T, et al. Neural networks-based regularization for large-scale medical image reconstruction. *Phys Med Biol*. 2020;65(13):135003.

155. Shen L, Pauly J, Xing L. NeRP: implicit neural representation learning with prior embedding for sparsely sampled image reconstruction. *IEEE Trans Neural Networks Learn Syst*. 2022.

156. Zhu B, Liu JZ, Cauley SF, Rosen BR, Rosen MS. Image reconstruction by domain-transform manifold learning. *Nature*. 2018;555(7697):487-492.

157. Eo T, Shin H, Jun Y, Kim T, Hwang D. Accelerating Cartesian MRI by domain-transform manifold learning in phase-encoding direction. *Med Image Anal*. 2020;63:101689.

158. Duan C, Deng H, Xiao S, et al. Fast and accurate reconstruction of human lung gas MRI with deep learning. *Magn Reson Med*. 2019;82(6):2273-2285.

159. Bilgic B, Chatnuntawech I, Manhard MK, et al. Highly accelerated multishot echo planar imaging through synergistic machine learning and joint reconstruction. *Magn Reson Med*. 2019;82(4):1343-1358.

160. Pramanik A, Aggarwal HK, Jacob M. Deep generalization of structured low-rank algorithms (Deep-SLR). *IEEE Trans Med Imaging*. 2020;39(12):4186-4197.

161. Souza R, Beauferris Y, Loos W, Lebel RM, Frayne R. Enhanced deep-learning-based magnetic resonance image reconstruction by leveraging prior subject-specific brain imaging: Proof-of-concept using a cohort of presumed normal subjects. *IEEE J Sel Top Signal Process*. 2020;14(6):1126-1136.

162. Quan TM, Nguyen-Duc T, Jeong WK. Compressed sensing MRI reconstruction using a generative adversarial network with a cyclic loss. *IEEE Trans Med Imaging*. 2018;37(6):1488-1497.

163. Quan C, Zhou J, Zhu Y, et al. Homotopic gradients of generative density priors for MR image reconstruction. *IEEE Trans Med Imaging*. 2021;40(12):3265-3278.

164. Yaqub M, Jinchao F, Ahmed S, et al. Gan-tl: Generative adversarial networks with transfer learning for mri reconstruction. *Appl Sci*. 2022;12(17):8841.

165. Wu Y, Ma Y, Capaldi D Pietro, et al. Incorporating prior knowledge via volumetric deep residual network to optimize the reconstruction of sparsely sampled MRI. *Magn Reson Imaging*. 2020;66:93-103.

166. Aghabiglou A, Eksioglu EM. MR image reconstruction using densely connected residual convolutional networks. *Comput Biol Med*. 2021;139:105010.

167. Du T, Zhang H, Li Y, et al. Adaptive convolutional neural networks for accelerating magnetic resonance imaging via k-space data interpolation. *Med Image Anal*. 2021;72:102098.

168. Wu Y, Ma Y, Liu J, Du J, Xing L. Self-attention convolutional neural network for improved MR image reconstruction. *Inf Sci (Ny)*. 2019;490:317-328.

169 Huang J, Fang Y, Wu Y, et al. Swin transformer for fast MRI. *Neurocomputing*. 2022;493:281-304.

170. Lin K, Heckel R. Vision transformers enable fast and robust accelerated mri. In: *International Conference on Medical Imaging with Deep Learning*. PMLR; 2022:774-795.

171. Bao L, Ye F, Cai C, et al. Undersampled MR image reconstruction using an enhanced recursive residual network. *J Magn Reson*. 2019;305:232-246.

172. Lønning K, Putzky P, Sonke JJ, Reneman L, Caan MWA, Welling M. Recurrent inference machines for reconstructing heterogeneous MRI data. *Med Image Anal*. 2019;53:64-78.

173. Yang G, Yu S, Dong H, et al. DAGAN: deep de-aliasing generative adversarial networks for fast compressed sensing MRI reconstruction. *IEEE Trans Med Imaging*. 2017;37(6):1310-1321.

174. Oh C, Kim D, Chung J, Han Y, Park H. A k-space-to-image reconstruction network for MRI using recurrent neural network. *Med Phys*. 2021;48(1):193-203.

175. Jalal A, Arvinte M, Daras G, Price E, Dimakis AG, Tamir J. Robust compressed sensing mri with deep generative priors. *Adv Neural Inf Process Syst*. 2021;34:14938-14954.

176. Peng C, Guo P, Zhou SK, Patel VM, Chellappa R. Towards performant and reliable undersampled MR reconstruction via diffusion model sampling. In: *International Conference on Medical Image Computing and Computer-Assisted Intervention*. Springer; 2022:623-633.

177. Luo G, Blumenthal M, Heide M, Uecker M. Bayesian MRI reconstruction with joint uncertainty estimation using diffusion models. *Magn Reson Med*. 2023;90(1):295-311.

178. Güngör A, Dar SUH, Öztürk Ş, et al. Adaptive diffusion priors for accelerated MRI reconstruction. *Med Image Anal*. 2023:102872.

179. Mardani M, Gong E, Cheng JY, et al. Deep generative adversarial neural networks for compressive sensing MRI. *IEEE Trans Med Imaging*. 2018;38(1):167-179.

180. Bhadra S, Zhou W, Anastasio MA. Medical image reconstruction with image-adaptive priors learned by use of generative adversarial networks. In: *Medical Imaging 2020: Physics of Medical Imaging*. Vol 11312. SPIE; 2020:206-213.

181. Shaul R, David I, Shitrit O, Raviv TR. Subsampled brain MRI reconstruction by generative adversarial neural networks. *Med Image Anal*. 2020;65:101747.

182. Yuan Z, Jiang M, Wang Y, et al. SARA-GAN: Self-attention and relative average discriminator based generative adversarial networks for fast compressed sensing MRI reconstruction. *Front Neuroinform*. 2020;14:611666.

183. Li Y, Li J, Ma F, Du S, Liu Y. High quality and fast compressed sensing MRI reconstruction via edge-enhanced dual discriminator generative adversarial network. *Magn Reson Imaging*. 2021;77:124-136.

184. Chen Y, Firmin D, Yang G. Wavelet improved GAN for MRI reconstruction. In: *Medical Imaging 2021: Physics of Medical Imaging*. Vol 11595. SPIE; 2021:285-295.

185. Kelkar VA, Bhadra S, Anastasio MA. Compressible latent-space invertible networks for generative model-constrained image reconstruction. *IEEE Trans Comput Imaging*. 2021;7:209-223.

186. Schlemper J, Castro DC, Bai W, et al. Bayesian deep learning for accelerated MR image reconstruction. In: *Machine Learning for Medical Image Reconstruction: First International Workshop, MLMIR 2018, Held in Conjunction with MICCAI 2018, Granada, Spain, September 16, 2018, Proceedings 1*. Springer; 2018:64-71.

187. Narnhofer D, Effland A, Kobler E, Hammernik K, Knoll F, Pock T. Bayesian uncertainty estimation of learned variational MRI reconstruction. *IEEE Trans Med Imaging*. 2021;41(2):279-291.

188. Luo G, Zhao N, Jiang W, Hui ES, Cao P. MRI reconstruction using deep Bayesian estimation. *Magn Reson Med*. 2020;84(4):2246-2261.

189. Desai AD, Ozturkler BM, Sandino CM, et al. Noise2Recon: Enabling SNR-robust MRI reconstruction with semi-supervised and self-supervised learning. *Magn Reson Med*. 2023.

190. Mardani LLM. Semi-supervised super-resolution gans for mri. In: *31st Conference on Neural Information Processing Systems (NIPS 2017), Long Beach, CA, USA*. ; 2017.

191. Wang Z, Lin Y, Cheng KTT, Yang X. Semi-supervised mp-MRI data synthesis with StitchLayer and auxiliary distance maximization. *Med Image Anal*. 2020;59:101565.

192. Desai AD, Gunel B, Ozturkler BM, et al. VORTEX: Physics-Driven Data Augmentations Using Consistency Training for Robust Accelerated MRI Reconstruction. *arXiv Prepr arXiv211102549*. 2021.

193. Yurt M, Dalmaz O, Dar S, et al. Semi-supervised learning of MRI synthesis without fully-sampled ground truths. *IEEE Trans Med Imaging*. 2022;41(12):3895-3906.

194. Nguyen H, Luo S, Ramos F. Semi-supervised Learning Approach to Generate Neuroimaging Modalities with Adversarial Training. In: *Advances in Knowledge Discovery and Data Mining: 24th Pacific-Asia Conference, PAKDD 2020, Singapore, May 11–14, 2020, Proceedings, Part II 24*. Springer; 2020:409-421.

195. Xiang L, Chen Y, Chang W, et al. Deep-learning-based multi-modal fusion for fast MR reconstruction. *IEEE Trans Biomed Eng*. 2018;66(7):2105-2114.

196. Gong K, Han P, El Fakhri G, Ma C, Li Q. Arterial spin labeling MR image denoising and reconstruction using unsupervised deep learning. *NMR Biomed*. 2022;35(4):e4224.

197. Yaman B, Hosseini SAH, Moeller S, Ellermann J, Uğurbil K, Akçakaya M. Self-supervised learning of

physics-guided reconstruction neural networks without fully sampled reference data. *Magn Reson Med*. 2020;84(6):3172-3191.

198. Hu C, Li C, Wang H, Liu Q, Zheng H, Wang S. Self-supervised learning for mri reconstruction with a parallel network training framework. In: *International Conference on Medical Image Computing and Computer-Assisted Intervention*. Springer; 2021:382-391.

199. Akçakaya M, Moeller S, Weingärtner S, Uğurbil K. Scan-specific robust artificial-neural-networks for k-space interpolation (RAKI) reconstruction: Database-free deep learning for fast imaging. *Magn Reson Med*. 2019;81(1):439-453.

200. Yaman B, Hosseini SAH, Moeller S, Ellermann J, Uğurbil K, Akçakaya M. Ground-truth free multi-mask self-supervised physics-guided deep learning in highly accelerated MRI. In: *2021 IEEE 18th International Symposium on Biomedical Imaging (ISBI)*. IEEE; 2021:1850-1854.

201. Yaman B, Shenoy C, Deng Z, et al. Self-supervised physics-guided deep learning reconstruction for high-resolution 3D LGE CMR. In: *2021 IEEE 18th International Symposium on Biomedical Imaging (ISBI)*. IEEE; 2021:100-104.

202. Yaman B, Gu H, Hosseini SAH, et al. Multi-mask self-supervised learning for physics-guided neural networks in highly accelerated magnetic resonance imaging. *NMR Biomed*. 2022;35(12):e4798.

203. Kustner T, Pan J, Gilliam C, et al. Self-supervised motion-corrected image reconstruction network for 4D magnetic resonance imaging of the body trunk. *APSIPA Trans Signal Inf Process*. 2022;11(1):e12.

204. Zhou B, Schlemper J, Dey N, et al. Dual-domain self-supervised learning for accelerated non-cartesian mri reconstruction. *Med Image Anal*. 2022;81:102538.

205. Munoz C, Qi H, Cruz G, Küstner T, Botnar RM, Prieto C. Self-supervised learning-based diffeomorphic non-rigid motion estimation for fast motion-compensated coronary MR angiography. *Magn Reson Imaging*. 2022;85:10-18.

206. Wang S, Wu R, Li C, et al. PARCEL: Physics-based Unsupervised Contrastive Representation Learning for Multi-coil MR Imaging. *IEEE/ACM Trans Comput Biol Bioinforma*. 2022.

207. Zou J, Li C, Jia S, et al. SelfCoLearn: Self-supervised collaborative learning for accelerating dynamic MR imaging. *Bioengineering*. 2022;9(11):650.

208. Pang T, Quan Y, Ji H. Self-supervised bayesian deep learning for image recovery with applications to compressive sensing. In: *Computer Vision–ECCV 2020: 16th European Conference, Glasgow, UK, August 23–28, 2020, Proceedings, Part XI 16*. Springer; 2020:475-491.

209. Aggarwal HK, Pramanik A, John M, Jacob M. ENSURE: A general approach for unsupervised training of deep image reconstruction algorithms. *IEEE Trans Med Imaging*. 2022;42(4):1133-1144.

210. Jung W, Lee HS, Seo M, et al. MR-self Noise2Noise: self-supervised deep learning–based image quality improvement of submillimeter resolution 3D MR images. *Eur Radiol*. 2023;33(4):2686-2698.

211. Millard C, Chiew M. A theoretical framework for self-supervised MR image reconstruction using sub-sampling via variable density Noisier2Noise. *IEEE Trans Comput imaging*. 2023.

212. Miller Z, Johnson KM. Motion compensated self supervised deep learning for highly accelerated 3D ultrashort Echo time pulmonary MRI. *Magn Reson Med*. 2023;89(6):2361-2375.

213. Xia Y, Ravikumar N, Greenwood JP, Neubauer S, Petersen SE, Frangi AF. Super-resolution of cardiac MR cine imaging using conditional GANs and unsupervised transfer learning. *Med Image Anal*. 2021;71:102037.

214. Lei K, Mardani M, Pauly JM, Vasanawala SS. Wasserstein GANs for MR imaging: from paired to unpaired training. *IEEE Trans Med Imaging*. 2020;40(1):105-115.

215. Sim B, Oh G, Kim J, Jung C, Ye JC. Optimal transport driven CycleGAN for unsupervised learning in inverse problems. *SIAM J Imaging Sci*. 2020;13(4):2281-2306.

216 Oh G, Sim B, Chung H, Sunwoo L, Ye JC. Unpaired deep learning for accelerated MRI using optimal transport driven CycleGAN. *IEEE Trans Comput Imaging*. 2020;6:1285-1296.

217. Chung H, Cha E, Sunwoo L, Ye JC. Two-stage deep learning for accelerated 3D time-of-flight MRA without matched training data. *Med Image Anal*. 2021;71:102047.

218. Oh G, Lee JE, Ye JC. Unpaired MR motion artifact deep learning using outlier-rejecting bootstrap aggregation. *IEEE Trans Med Imaging*. 2021;40(11):3125-3139.

219. Cha E, Chung H, Kim EY, Ye JC. Unpaired training of deep learning tMRA for flexible spatio-temporal resolution. *IEEE Trans Med Imaging*. 2020;40(1):166-179.

220. Zou Q, Ahmed AH, Nagpal P, Kruger S, Jacob M. Deep generative storm model for dynamic imaging. In: *2021 IEEE 18th International Symposium on Biomedical Imaging (ISBI)*. IEEE; 2021:114-117.

221. Mani MP, Aggarwal HK, Ghosh S, Jacob M. Model-based deep learning for reconstruction of joint kq under-sampled high resolution diffusion MRI. In: *2020 IEEE 17th International Symposium on Biomedical Imaging (ISBI)*. IEEE; 2020:913-916.

222. Chung H, Ye JC. Score-based diffusion models for accelerated MRI. *Med Image Anal*. 2022;80:102479.

223. Tezcan KC, Baumgartner CF, Luechinger R, Pruessmann KP, Konukoglu E. MR image reconstruction using deep density priors. *IEEE Trans Med Imaging*. 2018;38(7):1633-1642.

224. Cole EK, Ong F, Vasanawala SS, Pauly JM. Fast unsupervised mri reconstruction without fully-sampled ground truth data using generative adversarial networks. In: *Proceedings of the IEEE/CVF International Conference on Computer Vision*. ; 2021:3988-3997.

225. Korkmaz Y, Dar SUH, Yurt M, Özbey M, Cukur T. Unsupervised MRI reconstruction via zero-shot learned adversarial transformers. *IEEE Trans Med Imaging*. 2022.

226. Ahmed AH, Aggarwal H, Nagpal P, Jacob M. Dynamic MRI using deep manifold self-learning. In: *2020 IEEE 17th International Symposium on Biomedical Imaging (ISBI)*. IEEE; 2020:1052-1055.

227. Zou Q, Ahmed AH, Nagpal P, Priya S, Schulte RF, Jacob M. Variational manifold learning from incomplete data: application to multislice dynamic MRI. *IEEE Trans Med Imaging*. 2022;41(12):3552-3561.

228. Zou Q, Torres LA, Fain SB, Higano NS, Bates AJ, Jacob M. Dynamic imaging using motion-compensated smoothness regularization on manifolds (MoCo-SToRM). *Phys Med Biol*. 2022;67(14):144001.

229. Eun D in, Jang R, Ha WS, Lee H, Jung SC, Kim N. Deep-learning-based image quality enhancement of compressed sensing magnetic resonance imaging of vessel wall: comparison of self-supervised and unsupervised approaches. *Sci Rep*. 2020;10(1):13950.

230. Zhao D, Zhao F, Gan Y. Reference-driven compressed sensing MR image reconstruction using deep convolutional neural networks without pre-training. *Sensors*. 2020;20(1):308.

231. Liu J, Sun Y, Eldeniz C, Gan W, An H, Kamilov US. RARE: Image reconstruction using deep priors learned without groundtruth. *IEEE J Sel Top Signal Process*. 2020;14(6):1088-1099.

232. Yoo J, Jin KH, Gupta H, Yerly J, Stuber M, Unser M. Time-dependent deep image prior for dynamic

MRI. *IEEE Trans Med Imaging*. 2021;40(12):3337-3348.

233. Korkmaz Y, Yurt M, Dar SUH, Özbey M, Cukur T. Deep MRI reconstruction with generative vision transformers. In: *Machine Learning for Medical Image Reconstruction: 4th International Workshop, MLMIR 2021, Held in Conjunction with MICCAI 2021, Strasbourg, France, October 1, 2021, Proceedings 4*. Springer; 2021:54-64.

234. Ahmed AH, Zou Q, Nagpal P, Jacob M. Dynamic imaging using deep bi-linear unsupervised representation (deblur). *IEEE Trans Med Imaging*. 2022;41(10):2693-2703.

235. Wang K, Tamir JI, De Goyeneche A, et al. High fidelity deep learning-based MRI reconstruction with instance-wise discriminative feature matching loss. *Magn Reson Med*. 2022;88(1):476-491.

236. Duff MAG, Simpson IJA, Ehrhardt MJ, Campbell NDF. VAEs with structured image covariance applied to compressed sensing MRI. *Phys Med Biol*. 2023;68(16):165008.

237. Cui ZX, Jia S, Cao C, et al. K-UNN: k-space interpolation with untrained neural network. *Med Image Anal*. 2023;88:102877.

238. Lebel RM. Performance characterization of a novel deep learning-based MR image reconstruction pipeline. *arXiv Prepr arXiv200806559*. 2020.

239. Pezzotti N, de Weerdt E, Yousefi S, et al. Adaptive-CS-Net: FastMRI with adaptive intelligence. *arXiv Prepr arXiv191212259*. 2019.

240. Behl N. Deep resolve—mobilizing the power of networks. *MAGNETOM Flash*. 2021;1:29-35.

241. Zhai R, Huang X, Zhao Y, et al. Intelligent incorporation of AI with model constraints for MRI acceleration. In: *Proceedings of the 29th Annual Meeting of ISMRM*. ; 2021.

242. Park JC, Park KJ, Park MY, Kim M, Kim JK. Fast T2-Weighted Imaging With Deep Learning-Based Reconstruction: Evaluation of Image Quality and Diagnostic Performance in Patients Undergoing Radical Prostatectomy. *J Magn Reson Imaging*. 2022;55(6):1735-1744.

243. Zochowski KC, Tan ET, Argentieri EC, et al. Improvement of peripheral nerve visualization using a deep learning-based MR reconstruction algorithm. *Magn Reson Imaging*. 2022;85:186-192.

244. Chan C C, Haldar J P. Local perturbation responses and checkerboard tests: Characterization tools for nonlinear MRI methods. *Magnetic Resonance in Medicine*, 2021, 86(4): 1873-1887.

245. Dziadosz M, Rizzo R, Kyathanahally S P, et al. Denoising single MR spectra by deep learning: Miracle or mirage?. *Magnetic resonance in medicine*, 2023.

246. Blau Y, Michaeli T. The perception-distortion tradeoff. *Proceedings of the IEEE conference on computer vision and pattern recognition*. 2018: 6228-6237.

247. Antun V, Renna F, Poon C, Adcock B, Hansen AC. On instabilities of deep learning in image reconstruction and the potential costs of AI. *Proc Natl Acad Sci*. 2020;117(48):30088-30095.

248. Gottschling N M, Antun V, Hansen A C, et al. The troublesome kernel—On hallucinations, no free lunches and the accuracy-stability trade-off in inverse problem. *arXiv Prepr arXiv2001.01258*. 2020.

249. Colbrook M J, Antun V, Hansen A C. The difficulty of computing stable and accurate neural networks: On the barriers of deep learning and Smale's 18$^{th}$ problem. *PNAS*, 2022, 119(12): e2107151119.

250. Block KT. Subtle pitfalls in the search for faster medical imaging. *Proc Natl Acad Sci*. 2022;119(17):e2203040119.

251. Muckley MJ, Riemenschneider B, Radmanesh A, et al. Results of the 2020 fastMRI challenge for machine learning MR image reconstruction. *IEEE Trans Med Imaging*. 2021;40(9):2306-2317.

252. Zhu Z, Zhu X, Ohliger MA, et al. Coil combination methods for multi-channel hyperpolarized 13C imaging data from human studies. *J Magn Reson*. 2019;301:73-79.

253. Gu H, Yaman B, Moeller S, Ellermann J, Ugurbil K, Akçakaya M. Revisiting ℓ 1-wavelet compressed-sensing MRI in the era of deep learning. *Proc Natl Acad Sci*. 2022;119(33):e2201062119.

254. Liu X, Wang J, Liu F, Zhou SK. Universal undersampled MRI reconstruction. In: *Medical Image Computing and Computer Assisted Intervention–MICCAI 2021: 24th International Conference, Strasbourg, France, September 27–October 1, 2021, Proceedings, Part VI 24*. Springer; 2021:211-221.

255. Feng CM, Yan Y, Wang S, Xu Y, Shao L, Fu H. Specificity-preserving federated learning for MR image reconstruction. *IEEE Trans Med Imaging*. 2022.

256. Elmas G, Dar SUH, Korkmaz Y, et al. Federated learning of generative image priors for MRI reconstruction. *IEEE Trans Med Imaging*. 2022.

257. Wu R, Li C, Zou J, Wang S. FedAutoMRI: Federated Neural Architecture Search for MR Image Reconstruction. *arXiv Prepr arXiv230711538*. 2023.

258. Zou J, Li C, Wu R, Pei T, Zheng H, Wang S. Self-Supervised Federated Learning for Fast MR Imaging. *arXiv Prepr arXiv230506066*. 2023.

259. Wang K, Kellman M, Sandino CM, et al. Memory-efficient learning for high-dimensional mri reconstruction. In: *Medical Image Computing and Computer Assisted Intervention–MICCAI 2021: 24th International Conference, Strasbourg, France, September 27–October 1, 2021, Proceedings, Part VI 24*. Springer; 2021:461-470.

260. Pramanik A, Jacob M. Accelerated parallel MRI using memory efficient and robust monotone operator learning (MOL). In: *2023 IEEE 20th International Symposium on Biomedical Imaging (ISBI)*. IEEE; 2023:1-4.

261. Mirza MU, Dalmaz O, Bedel HA, et al. Learning Fourier-Constrained Diffusion Bridges for MRI Reconstruction. *arXiv Prepr arXiv230801096*. 2023.

## Table captions

Table S1: Overview of various fast MRI techniques leveraging deep learning, highlighting their deep learning type, application domain, acceleration factor, sampling trajectory, data source, and code source

## Figure Captions

1. Figure 1. Illustration of the two underlying equations governing the MR imaging process: the Bloch equation and Signal equation. The Bloch equation creates clinical useful image contrasts by manipulating the radiofrequency fields. The signal equation establishes the discrete Fourier transform relationship between the acquired k-space samples and the final images. $\mathcal{M}$ denotes the physic model of the specific pulse sequence which can be analyzed by solving Bloch equation, while $K$, $F$ and $C$ denote the discrete sampling process, Fourier encoding process and parallel receiving coil sensitivity encoding process respectively.

2. Figure 2. The MR image reconstruction pipeline includes the primary step of inverse based reconstruction from incomplete k-space samples, and multiple raw data and image processing steps. Theoretically, the deep learning techniques can benefit the whole MR image reconstruction pipeline, however this review mainly focuses on the primary step of MR image reconstruction using deep learning and three common MR application scenarios including the contrast weighted, dynamic and quantitative imaging.

3. Figure 3. Illustration of supervised learning-based MRI reconstruction. $\ell(\cdot)$ is the objective function between the network output $f(m_\Omega^i, E_\Omega^i; \theta)$ parametrized by $\theta$ for subsampled k-space data $m_\Omega^i$ and the reference image $v^i$. $E_\Omega^i$ is the corresponding encoding matrix and $\Omega$ represents the subsampling patterns used.

4. Figure 4. Illustration of semi-supervised learning-based MRI reconstruction. $\ell 1(\cdot)$ is the supervised objective function between the network output $f(m_\Omega^i, E_\Omega^i; \theta)$ parametrized by $\theta$ for subsampled k-space data $m_\Omega^i$ and the labeled reference image $v^i$. $\ell 2(\cdot)$ is the unsupervised objective function between the unlabeled reference image $m^j$ and the pseudo-label $\hat{v}^j$ generated by the supervised training. $E_\Omega^i$ is the corresponding encoding matrix and $\Omega$ represents the subsampling patterns used.

5. Figure 5. Illustration of unsupervised learning-based MRI reconstruction, where the sub-sampled data indices $\Omega$ is divided into two sets $\Psi$ and $\Lambda$. $\ell(\cdot)$ is the objective function between the network output $f(m_\Psi^i, E_\Psi^i, \theta)$ parametrized by $\theta$ for subsampled k-space data $m_\Psi^i$ and $m_\Lambda^i$, the set of k-space point in $\Lambda$ used as reference image. $E_\Lambda^i$ represents the encoding operator specified by the k-space indices in $\Lambda$.

# Supporting information

Table S1: Overview of various fast MRI techniques leveraging deep learning, highlighting their deep learning type, application domain, acceleration factor, sampling trajectory, data and code source.

| Learning Type | Application Domain | Acceleration | Sampling Trajectory | Database | Anatomy | Code | Reference |
|---|---|---|---|---|---|---|---|
| | Contrast-weighted Imaging | 4.x | Cartesian | fastMRI | Knee | https://github.com/VLOGroup/mri-variationalnetwork | Hammernik et al. [1] |
| | Contrast-weighted Imaging, dynamic Imaging | 2.x, 8.x | Cartesian | - | Knee, cine images | https://github.com/MRSRL/complex-networks-release | Cole et al. [36] |
| | Contrast-weighted Imaging | 10.x | Non-Cartesian | - | Knee, Brain | https://github.com/guanhuaw/Bjork | Wang et al. [77] |
| | Contrast-weighted Imaging | 4.x, 5.x | Cartesian | fastMRI | Brain, Knee | https://github.com/pdawood/iterativeRaki | Dawood et al. [61] |
| | Contrast-weighted Imaging | 4.x, 6.x, 8.x, 10.x | Cartesian | fastMRI | Brain, Knee | https://github.com/hkaggarwal/J-MoDL | Aggarwal et al. [69] |
| | Contrast-weighted Imaging | 3.x, 4.x | Cartesian | ADNI | Brain | - | Eo et al. [21] |
| | Quantitative Imaging | 5.x, 8.x | Cartesian | - | Knee | - | Liu et al. [126] |
| | Contrast-weighted Imaging | 3.x, 5.x | Cartesian, Non-Cartesian | - | Knee, Liver | - | Liu et al. [75] |
| | Contrast-weighted Imaging | 4.x, 8.x | Cartesian, Non-Cartesian | SIAT MRI dataset, fastMRI | Brain, Knee | https://github.com/yqx7150/EDMSPRec | Zhang et al. [149] |
| | Contrast-weighted Imaging | - | Non-Cartesian | - | Lung, Liver, Abdomen | | Freedman et al. [109] |

| Learning Type | Application Domain | Acceleration | Sampling Trajectory | Database | Anatomy | Code | Reference |
|---|---|---|---|---|---|---|---|
| Supervised | Dynamic Imaging | 4.x, 12.x | Cartesian | Cardiac cine MRI | Cardiac | https://github.com/js3611/Deep-MRI-Reconstruction | Schlemper et al. [97] |
| | Contrast-weighted Imaging | 10.x | Cartesian | MRI data | Brain | https://github.com/hkaggarwal/modl | Aggarwal et al. [4] |
| | Contrast-weighted Imaging | 10.x, 3.3.x, 5.x | Cartesian, Non-Cartesian | MICCAI 2013 grand challenge data, Bowl cardiac Challenge | Brain, Cardiac | https://github.com/lonelyatu/TRANS-Net | Hu et al. [142] |
| | Contrast-weighted Imaging | 4.x | Cartesian | fastMRI, IXI data, Synthetic data | Brain, Knee | - | Wang et al. [82] |
| | Dynamic Imaging | 6.x | Cartesian | In-house | Cardiac | https://github.com/Keziwen/DIMENSION | Wang et al. [23] |
| | Contrast-weighted Imaging | 3.x-6.7.x | Cartesian, Non-Cartesian | SRI24-spm8 and Brainweb | Brain | https://github.com/yqx7150 | Liu et al. [90] |
| | Contrast-weighted Imaging | 3.x, 5.x | Cartesian | In-house | Brain | - | Wang et al. [2] |
| | Contrast-weighted Imaging | 2.x, 3.x, 4.x, 5.x | Non-Cartesian | the 2016 AAPM low-dose CT Grand Challenge, HCP | Brain, Abdomen | - | Han et al. [3] |
| | Contrast-weighted Imaging | 2.5x;3.3x;5x | Cartesian, Non-Cartesian | In-house | Brain | https://github.com/yangyan92/Pytorch_ADMM-CSNet | Yang et al. [5] |
| | Contrast-weighted Imaging | 3.x, 4.x, 5.x, 10.x | Cartesian | Coronal Spin Density Weighted without Fat Suppression, private | Brain, Knee | https://github.com/CedricChing/DeepMRI | Wang et al. [37] |
| | Contrast-weighted Imaging | 4.x | Cartesian, Non-Cartesian | fastMRI, MRI data | Knee, Brain | https://github.com/mrirecon/deep-deep-learning-with-bart | Blumenthal et al. [17] |
| | Contrast-weighted Imaging | 12.x, 16.x | Cartesian | http://mridata.org/ | Knee | https://github.com/ad12/meddlr | Desai et al. [189] |
| | Contrast-weighted Imaging | 3.x | N/A | Stanford Knee MRI data | Knee | https://github.com/gongenhao/GANCS.git | Mardani et al. [190] |

| Learning Type | Application Domain | Acceleration | Sampling Trajectory | Database | Anatomy | Code | Reference |
|---|---|---|---|---|---|---|---|
| **Semi-supervised** | Contrast-weighted Imaging | 16.x, 8.x | Cartesian | http://mridata.org/, fastMRI | Brain | https://github.com/ad12/meddlr | Desai et al. [192] |
| | Contrast-weighted Imaging | 2.x-10.x; 2.x-4.x | Cartesian | IXI dataset, In vivo brain dataset | Brain | https://github.com/icon-lab/ssGAN | Yurt et al. [193] |
| | | N/A | N/A | BraTS2020 | | https://github.com/Huangziqi777/DS-3_Net | Huang et al. [18] |
| **Un-supervised** | Contrast-weighted Imaging | 5x;6.6x;10x | Non-Cartesian | In-house | whole-body | - | Liu et al. [231] |
| | Contrast-weighted Imaging | 8.x, 2.x | Cartesian | 3D knee data | Knee | - | Yaman et al. [202] |
| | Contrast-weighted Imaging | 2.x, 4.x | Non-Cartesian | Human Connectome Project (HCP) | Brain | - | Zhou et al. [204] |
| | Contrast-weighted Imaging | 2.x, 4.x | Cartesian | fastMRI | Brain, Knee | - | Aggarwal et al. [209] |
| | Contrast-weighted Imaging | 4.x, 8.x | Cartesian | fastMRI | Brain and knee | https://github.com/charlesmillard/Noisier2Noise_for_recon | Millard et al. [211] |
| | Contrast-weighted Imaging | 2.x, 4.x, 6.x, 8.x | Cartesian | fastMRI, private | Brain, Knee | https://github.com/byaman14/SSDU | Yaman et al. [197] |
| | Contrast-weighted Imaging | 4.x | Cartesian | fastMRI | Knee | - | Sim et al. [215] |
| | Dynamic Imaging, Contrast-weighted Imaging | 5.x | Cartesian | - | Knee, Abdomen | https://github.com/MRSRL/unsupGAN-release | Cole et al. [224] |
| | contrast-weighted imaging | 13.93.x | Cartesian | In-house | Angiography | - | Cha et al. [219] |
| | Dynamic Imaging | N/A | Non-Cartesian | - | Pulmonary | - | Zou et al. [228] |
| | Dynamic Imaging | N/A | N/A | Cardiac cine data | Cardiac | - | Ahmed et al. [234] |
| | contrast-weighted imaging | 2x;3.3x;5x | Cartesian, Non-Cartesian | fastMRI | Knee | - | Duff et al. [236] |

| Learning Type | Application Domain | Acceleration | Sampling Trajectory | Database | Anatomy | Code | Reference |
|---|---|---|---|---|---|---|---|
| | contrast-weighted imaging | 3.x, 4.x, 8.x | Cartesian | fastMRI | | https://github.com/ternencewu123/PARCEL | Wang et al.[206] |